\documentclass[lettersize,journal]{IEEEtran}
\usepackage{amsmath,amsfonts}
\usepackage{algorithmic}
\usepackage{array}
\usepackage{subfigure}
\usepackage{xcolor} 
\usepackage{textcomp}
\usepackage{stfloats}
\usepackage{url}
\usepackage{verbatim}
\usepackage{multirow}
\usepackage{amsmath}
\usepackage{threeparttable}
\usepackage{tabularray}
\usepackage{graphicx}
\usepackage{booktabs}
\def\BibTeX{{\rm B\kern-.05em{\sc i\kern-.025em b}\kern-.08em
    T\kern-.1667em\lower.7ex\hbox{E}\kern-.125emX}}
\usepackage{balance}
\begin{document}
\title{PromptCast: A New Prompt-based Learning Paradigm for Time Series Forecasting}
\author{Hao Xue, Flora D. Salim \\
School of Computer Science and Engineering, University of New South Wales, Sydney, Australia
\thanks{Hao Xue and Flora D. Salim are with the School of Computer Science and Engineering, UNSW Sydney, hao.xue1@unsw.edu.au, flora.salim@unsw.edu.au}}

\newcommand{\name}{PISA}
\newcommand{\tname}{PromptCast}
\newcommand{\ie}{\textit{i.e.}}
\newcommand{\eg}{\textit{e.g.}}
\newcommand{\datasheet}[1]{{\color{blue}{{#1}}}}
\newcommand{\blue}[1]{{\color{blue}{{#1}}}}
\newcommand{\red}[1]{{\color{red}{{#1}}}}

% \markboth{Journal of \LaTeX\ Class Files,~Vol.~18, No.~9, September~2020}%
% {How to Use the IEEEtran \LaTeX \ Templates}

\maketitle

\begin{abstract}
This paper presents a new perspective on time series forecasting. In existing time series forecasting methods, the models take a sequence of numerical values as input and yield numerical values as output. The existing SOTA models are largely based on the Transformer architecture, modified with multiple encoding mechanisms to incorporate the context and semantics around the historical data. Inspired by the successes of pre-trained language foundation models, we pose a question about whether these models can also be adapted to solve time-series forecasting. Thus, we propose a new forecasting paradigm: prompt-based time series forecasting (PromptCast). In this novel task, the numerical input and output are transformed into prompts and the forecasting task is framed in a sentence-to-sentence manner, making it possible to directly apply language models for forecasting purposes. To support and facilitate the research of this task, we also present a large-scale dataset (PISA) that includes three real-world forecasting scenarios. We evaluate different SOTA numerical-based forecasting methods and language generation models. The benchmark results with various forecasting settings demonstrate the proposed PromptCast with language generation models is a promising research direction. Additionally, in comparison to conventional numerical-based forecasting, PromptCast shows a much better generalization ability under the zero-shot setting. 
\end{abstract}

\begin{IEEEkeywords}
time series forecasting, natural language generation, dataset and benchmark
\end{IEEEkeywords}

\section{Introduction}
\begin{figure*}
    \centering
    \includegraphics[width=0.7\textwidth]{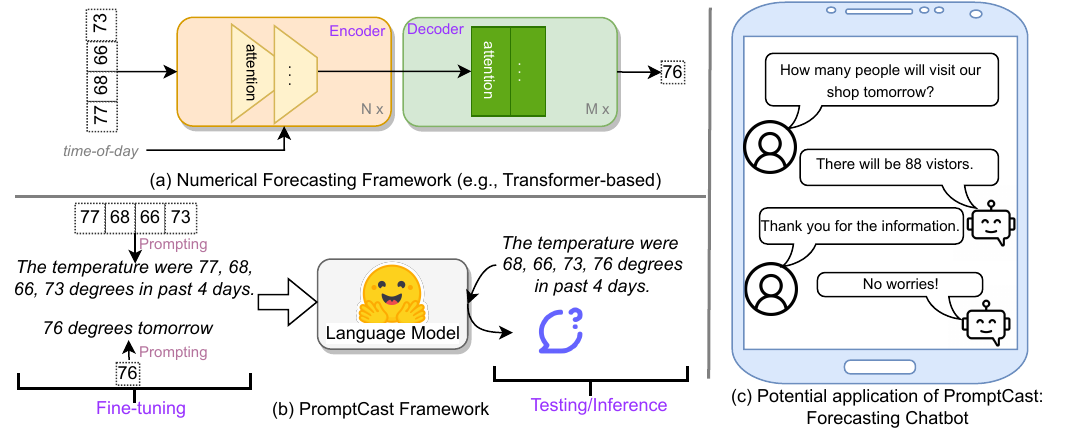}
    \caption{Conceptual illustrations of: (a) the typical framework of existing numerical-based forecasting method; (b) the framework of the proposed \tname; (c) a potential forecasting chatbot application based on \tname.}
    \label{fig:intro}
    % \vspace{-2ex}
\end{figure*}

Time series forecasting is a research-intensive field, especially with the increasing of applying various deep learning frameworks for prediction such as models based on LSTM~\cite{hochreiter1997long}, Temporal Convolutional Network (TCN)~\cite{lea2017temporal}, and Transformer~\cite{vaswani2017attention}.
More recently, we are witnessing a fast growth of large-scale pre-trained models in the Natural Language Processing (NLP) field.
These models, also known as foundation models~\cite{bommasani2021opportunities}, are often pre-trained with an extremely large amount of data and have demonstrated good performance across various downstream tasks. For example, BERT~\cite{bert} can be adapted for multiple NLP tasks, CLIP~\cite{clip} and GLIP~\cite{glip} are good at CV tasks.
However, we also notice that this evolution seems mostly limited to the NLP and CV fields.
Hence, we are particularly interested in exploring the research question of whether we can take advantage of large-scale pre-trained foundation models and adapt these models for predicting time series. 
To investigate this question, in this paper, we formally introduce a novel task: prompt-based time series forecasting (\tname).
The existing forecasting methods including the state-of-the-art Transformer-based forecasting models~\cite{li2019enhancing,zhou2021informer,liu2021pyraformer,xu2021autoformer,FEDformer,TACTIS} can be simplified as a numerical forecasting paradigm as shown in Figure~\ref{fig:intro} (a). Numerical forecasting methods always take numerical values as input and generate numerical values as the prediction for the next time step.
Instead, the input and output of the proposed prompt-based forecasting (Figure~\ref{fig:intro} (b)) are natural language sentences. This paradigm change enables the utilization of language generation models for forecasting.

This new forecasting paradigm is beneficial in multiple aspects.
% Forecasting under the \tname\ setting could enhance the generalization ability of forecasting. The pre-training techniques are well progressed for language models. Such techniques could also be applied to strengthen the generalization if we directly use language models for forecasting.
\tname\ presents a novel ``code less'' solution for time series forecasting, which could provide a new perspective rather than purely focusing on designing more and more complicated deep learning forecasting models (\eg, Transformer-based Informer, Autoformer, and FEDformer).
It also becomes a relatively easy-accessible and user-friendly method for non-researcher users, compared to existing forecasting models that require many tedious parameter searching and training processes, especially under a new forecasting scenario. 
As pointed out in a recent research~\cite{reed2022generalist}, the benefits of using a single neural model across different tasks are significant. The \tname\ task explores the potential of using language foundation models for the forecasting task which could make it possible to broaden the language models beyond the realm of typical text-based tasks.
In addition, it could inspire new research directions and new applications to better serve society. For example, as illustrated in Figure~\ref{fig:intro} (c), a chatbot with forecasting ability is one of the prospective future applications driven by the research of \tname\ task.
Currently, although AI-powered intelligent assistants or chatbots like Siri and Alexa can answer queries about general topics, they still fail to answer specific time series forecasting questions. With the help of \tname\ related research, they would be able to yield predictions based on the given contexts. 

To the best of our knowledge, this is the first effort in approaching general time-series forecasting from a language-based perspective without any modifications on the model architectures, resulting also in the first large-scale dataset, \name\ (\textbf{P}rompt based t\textbf{I}me \textbf{S}eries forec\textbf{A}sting), tailored for the task of prompt-based time series forecasting. It covers three real-world forecasting scenarios: weather temperature forecasting, energy consumption forecasting, and customer flow forecasting.
We believe that the release of this dataset will not only support the research of the \tname\ task but also have a great potential to stimulate related research in the time series analysis domain.
We further develop a benchmark in which we report the forecasting performance of multiple methods including both numerical-based forecasting methods and language generation models. To evaluate the generalization ability of \tname, this benchmark also explores various forecasting settings such as train-from-scratch, zero-shot prediction, multi-step forecasting, and multivariate forecasting.
In summary, our contributions are three-fold: 
\textbf{(1)} We propose a novel prompt-based forecasting paradigm, which differs from the existing forecasting methods. This is the first time the general time series forecasting problem is addressed in a natural language generation manner.
\textbf{(2)} We release a large-scale dataset (\name) with 311,932 data instances in total for the newly introduced task. The dataset covers diverse time series forecasting scenarios.
\textbf{(3)} We develop a benchmark (our data and codes for the benchmark are available at \url{https://github.com/HaoUNSW/PISA}) on the proposed \name\ datasets. It evaluates the state-of-the-art numerical-based forecasting methods and popular language generation models.

\section{Prompt-Based Time Series Forecasting}
The prompt-based time series forecasting task is developed from the general time series task. Here we first describe the general numerical-based forecasting and then formulate the proposed \tname\ task.
Let $\mathcal{U}= \{U_1 ,U_2, \cdots, U_M\}$ denotes a set of $M$ \textit{objects-of-interest}. Depending on different specific forecasting scenarios, the objects-of-interest could stand for different objects. For example, the objects could be places-of-interest (POI) such as bars and parks in human mobility forecasting~\cite{xue2022translating} or cities in weather forecasting.
% Under the general numerical time series forecasting task setting, the input is a history records of interested numerical data points collected on $n$ continuous time steps (\eg, daily data): $x^m_{t_1: t_{\text{obs}}}=[x^m_{t_1}, x^m_{t_2}, \cdots, x^m_{t_{\text{obs}}}~|~x^m_t \in \mathbb{R}^d]$, where $x^m_t$ represents the value of object-of-interest $U_m$ observed on time step $t$. 
% The forecasting target (output) is the numerical data value $x^{m}_{t_{{\text{obs}}+1}}$ of the next time step $t_{{\text{obs}}+1}$.
% Especially, the task could be univariate time series forecasting ($d=1$) or multivariate time series forecasting ($d>1$). Note that although we focus on the univariate time series to introduce the novel prompt-based time series forecasting task in this work, the proposed \tname\ can also be easily applied in the multivariate time series setting.
Under the general numerical time series forecasting task setting, the input is a historical record of interested numerical data points collected on $t_{\text{obs}}$ continuous time steps (\eg, daily data): $x^m_{t_1: t_{\text{obs}}}=[x^m_{t_1}, x^m_{t_2}, \cdots, x^m_{t_{\text{obs}}}~|~x^m_t]$, where $x^m_t$ represents the value of object-of-interest $U_m$ observed on time step $t$. 
The forecasting target (output) is the numerical data value $x^{m}_{t_{{\text{obs}}+1}}, x^{m}_{t_{{\text{obs}}+2}}, \cdots, x^{m}_{t_{{\text{obs}}+n}}$ of the next $n$ time steps $t_{{\text{obs}}+1}, t_{{\text{obs}}+2}, \cdots, t_{{\text{obs}}+n}$ ($n$ is the prediction horizon).

% Especially, the task could be univariate time series forecasting ($d=1$) or multivariate time series forecasting ($d>1$). Note that although we focus on the univariate time series to introduce the novel prompt-based time series forecasting task in this work, the proposed \tname\ can also be easily applied in the multivariate time series setting.

\begin{table*}[]
\centering
\caption{Templates for transforming \name-numerical to \name-prompt.}
\label{tab:dataset_template}
\scriptsize
\addtolength{\tabcolsep}{-0.65ex}
\begin{tabular}{l|c|c|p{2.0in}|p{3.0in}} \toprule
\multicolumn{3}{c|}{}  & Template & Example \\ \hline
\multirow{3}{*}{CT} & \multirow{2}{*}{\begin{tabular}[c]{@{}c@{}}Input Prompt\\ (Source)\end{tabular}} & Context & From \{$t_1$\} to \{$t_{\text{obs}}$\}, the average temperature of region \{$U_m$\} was \{$x^m_{t_1: t_{\text{obs}}}$\} degree on each day. & From August 16, 2019, Friday to August 30, 2019, Friday, the average temperature of region 110 was 78, 81, 83, 84, 84, 82, 83, 78, 77, 77, 74, 77, 78, 73, 76 degree on each day. \\ \cline{3-5}
 &  & Question & What is the temperature going to be on \{$t_{{\text{obs}}+1}$\}? & What is the temperature going to be on August 31, 2019, Saturday? \\ \cline{2-5}
 & \begin{tabular}[c]{@{}c@{}}Output Prompt\\  (Target)\end{tabular} & Answer & The temperature will be \{$x^{m}_{t_{{\text{obs}}+1}}$\} degree. & The temperature will be 78 degree. \\ \midrule
\multirow{3}{*}{ECL} & \multirow{2}{*}{\begin{tabular}[c]{@{}c@{}}Input Prompt\\ (Source)\end{tabular}} & Context & From \{$t_1$\} to \{$t_{\text{obs}}$\}, client \{$U_m$\} consumed \{$x^m_{t_1: t_{\text{obs}}}$\} kWh of electricity on each day. & From May 16, 2014, Friday to May 30, 2014, Friday, client 50 consumed 8975, 9158, 8786, 8205, 7693, 7419, 7595, 7596, 7936, 7646, 7808, 7736, 7913, 8074, 8329 kWh of electricity on each day. \\ \cline{3-5}
 &  & Question & What is the consumption going to be on \{$t_{{\text{obs}}+1}$\}? & What is the consumption going to be on May 31, 2014, Saturday? \\ \cline{2-5}
 & \begin{tabular}[c]{@{}c@{}}Output Prompt\\ (Target)\end{tabular} & Answer & This client will consume \{$x^{m}_{t_{{\text{obs}}+1}}$\} kWh of electricity. & This client will consume 8337 kWh of electricity. \\ \midrule
\multirow{3}{*}{SG} & \multirow{2}{*}{\begin{tabular}[c]{@{}c@{}}Input Prompt\\ (Source)\end{tabular}} & Context & From \{$t_1$\} to \{$t_{\text{obs}}$\}, there were \{$x^m_{t_1: t_{\text{obs}}}$\} people visiting POI \{$U_m$\} on each day. & From May 23, 2021, Sunday to June 06, 2021, Sunday, there were 13, 17, 13, 20, 16, 16, 17, 17, 19, 20, 12, 12, 14, 12, 13 people visiting POI 324 on each day. \\ \cline{3-5}
 &  & Question & How many people will visit POI \{$U_m$\} on \{$t_{{\text{obs}}+1}$\}? & How many people will visit POI 324 on June 07, 2021, Monday? \\ \cline{2-5}
 & \begin{tabular}[c]{@{}c@{}}Output Prompt\\ (Target)\end{tabular} & Answer & There will be \{$x^{m}_{t_{{\text{obs}}+1}}$\} visitors. & There will be 15 visitors. \\ \bottomrule
\end{tabular}
\end{table*}

The overarching goal of the \tname\ task is to leverage language foundation models to forecast time series in a sentence-to-sentence fashion.
To achieve this goal, based on the above formulated problem of numerical time series forecasting, the numerical values need to be transferred and described as natural language sentences. This data-to-text transformation is referred to as a prompting process in this work (the details of prompting are presented in the next section). Specifically, the input numerical sequence $x^m_{t_1: t_{\text{obs}}}$ is turned into input prompts, and the forecasting target values are transformed as the output prompt. Consequently, time series forecasting can be addressed through a natural language generation paradigm, and language foundation models can be adopted as the core forecasting models in \tname\ task.
For simplification, we primarily use the fundamental univariate time series single-step forecasting as the default setting to illustrate the concept of our novel \tname\ paradigm in the main paper.
More discussion and experimental results of \tname\ in multi-step forecasting and multivariate forecasting are reported in Sec.~\ref{sec:multi} and Sec.~\ref{sec:mts}.

\section{Dataset Design and Description}
In this section, we demonstrate the design and construction of the proposed \name\ dataset. The overall designing guideline is: (1) to preprocess original data given in the numerical format (raw data) for the forecasting task setting (Sec.~\ref{sec:raw_data}) and (2) to transform the numerical data to natural language input/output formats with prompts (Sec.~\ref{sec:prompt_data}). We also describe the features and statistics (Sec.~\ref{sec:data_sta}).

\subsection{Data Sources and Processing}\label{sec:raw_data}
To create a diverse dataset, we consider three real-world forecasting scenarios (3 sub-sets of our \name\ dataset) from various domains: weather forecasting, energy consumption forecasting, and human mobility forecasting. The data sources for these scenarios are:
% \begin{itemize}
% \setlength{\itemsep}{0pt}
% \setlength{\parskip}{0pt}
% \setlength{\parsep}{0pt}
    % \item 
    \textbf{City Temperature (CT)}: This data source\footnote{https://academic.udayton.edu/kissock/http/Weather/default.htm} provides the daily average temperature (in Fahrenheit degrees) of multiple cities globally. 110 international cities are randomly selected to form the dataset.
    % \item 
    \textbf{Electricity Consumption Load (ECL)}: The original data includes the electricity consumption values (in Kwh) of 321 users\footnote{https://archive.ics.uci.edu/ml/datasets/ElectricityLoadDiagrams20112014}. We eliminated users with incomplete records and randomly selected 50 users with complete data for the entire collection period. Additionally, the hourly usage values for each chosen user were consolidated into daily usage data.
    % \item 
    \textbf{SafeGraph Human Mobility Data (SG)}: This real-world human mobility data from SafeGraph Weekly Patterns\footnote{https://docs.safegraph.com/docs/weekly-patterns\#section-weekly-patterns-schema} contains the daily raw counts of visitors to POIs. We expanded the data collection from 5 months in~\cite{xue2022translating} to almost 15 months and then randomly selected 324 POIs with full records. 
% \end{itemize}
The exact data collection periods are reported in Table~\ref{tab:dataset}. Following the standard protocol~\cite{xu2021autoformer,xue2022translating}, each sub-set is divided into train/val/test at the ratio of 7:1:2 by the chronological order (Table~\ref{tab:dataset}).
The numerical sequence of each object-of-interest in each sub-set is then split into multiple instances (for training/validation/test) by applying sliding windows.
The window size equals $t_{\text{obs}} + 1$ (including $t_{\text{obs}}$ time steps as input historical data and 1 step as the forecasting target) and the step size of the sliding window is 1 day.
Specifically, following previous work~\cite{xue2022translating}, the observation length of the input sequence is set as 15 ($t_{\text{obs}}=15$).
To distinguish the numerical data used for numerical methods and the language-based dataset processed for language models, the numerical sequences processed by the above sliding window are referred to as \name-numerical whereas the other is named as \name-prompt (see next subsection).
% Note that the instances in \name-numerical one-to-one mapping

\noindent\textbf{Ethical Considerations.}
The only possible sensitive information is the identifier of the object-of-interest (\eg, the POI id in SG). To remove this, we randomly assigned object-of-interest index $U_m$ starting from 1 to $M$.
Given that the original data source for each subset is aggregate statistics with no personally identifiable information, the generated \name\ dataset does not contain any private information that can be deciphered.

\subsection{Template-Based Prompting}\label{sec:prompt_data}
The core of the proposed \tname\ task is shaping the time series forecasting in a language generation manner. To serve this purpose, a key step in building a dataset for \tname\ is to describe and transform the numerical sequential data (\ie, \name-numerical) to natural language sentences.
As demonstrated in~\cite{xue2022translating}, using template-based description is an effective and efficient approach to achieve the data-to-text transformation.
In this work, we explicitly introduce three templates for the three sub-sets and Table~\ref{tab:dataset_template} lists the templates and the corresponding examples.

In a nutshell, the template consists of two main parts: input prompt and output prompt.
The input prompt covers the description of the historical observation and the indicators of the prediction target time step (\ie, $t_{{\text{obs}} + 1}$). The output prompt handles the desired prediction value ($x^{m}_{t_{{\text{obs}}+1}}$) which is used as the ground truth label for training or evaluation. This input/output prompt setting is similar to the source/target sentence in machine translation.
For researchers who are more familiar with the open question-answering setting, our \name\ dataset can also be interpreted as a question-answering task setting.
The input prompt can be broken into the context part and the question part.
The context provides the historical information for forecasting and the question part can be seen as the input query about the future. Naturally, the output prompt is the ground truth answer that responds to the question.
Based on the templates and the processed numerical sequences, \name-prompt is then generated.
Note that the instances in \name-numerical map the instances in \name-prompt.
For example, the first instance in \name-prompt is transferred from the first instance in \name-numerical.
This is to ensure that they can be used to compare the performance of numerical forecasting methods and the language-based forecasting methods in the benchmark.
Our \name-prompt provides the input prompts and the corresponding output prompts in separate files (\eg, val\_x\_prompt.txt and val\_y\_prompt.txt).

\subsection{Statistics Overview}\label{sec:data_sta}
\begin{figure*}
    \centering
    \subfigure[City Temperature (CT)]{
    \includegraphics[width=.275\textwidth]{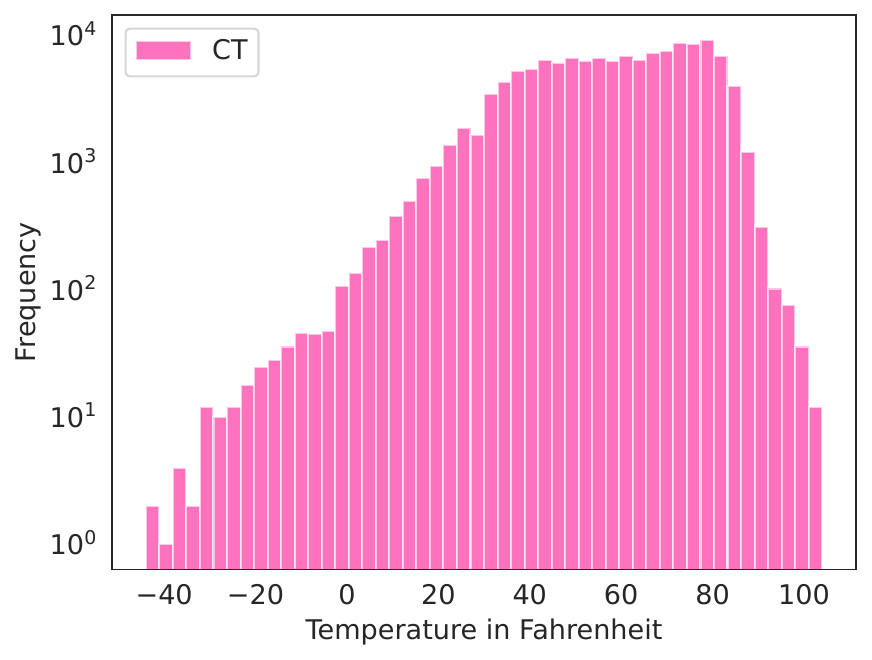}}
    \subfigure[Electricity Consumption Load (ECL)]{
    \includegraphics[width=.275\textwidth]{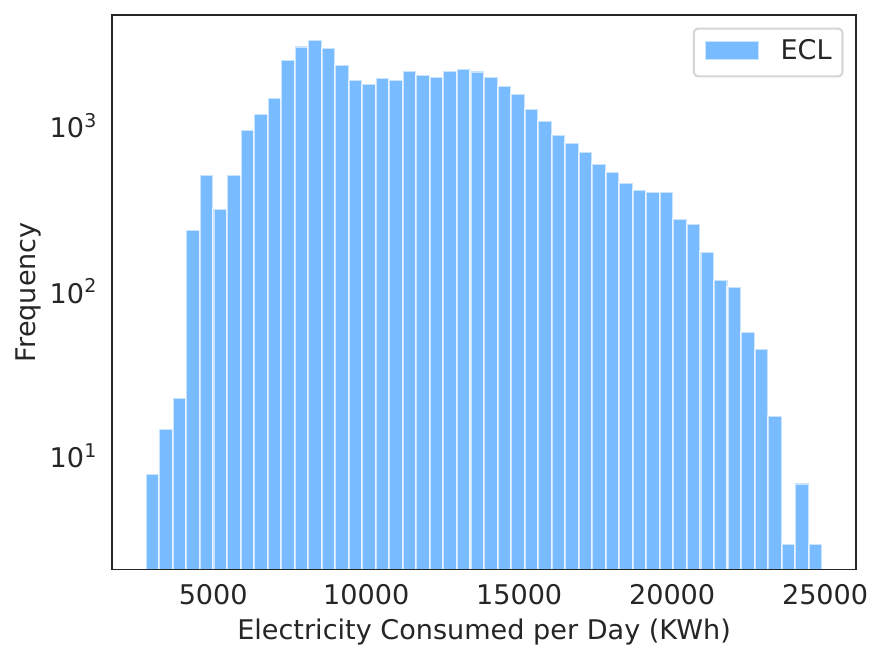}} 
    \subfigure[SafeGraph Human Mobility Data (SG)]{
    \includegraphics[width=.275\textwidth]{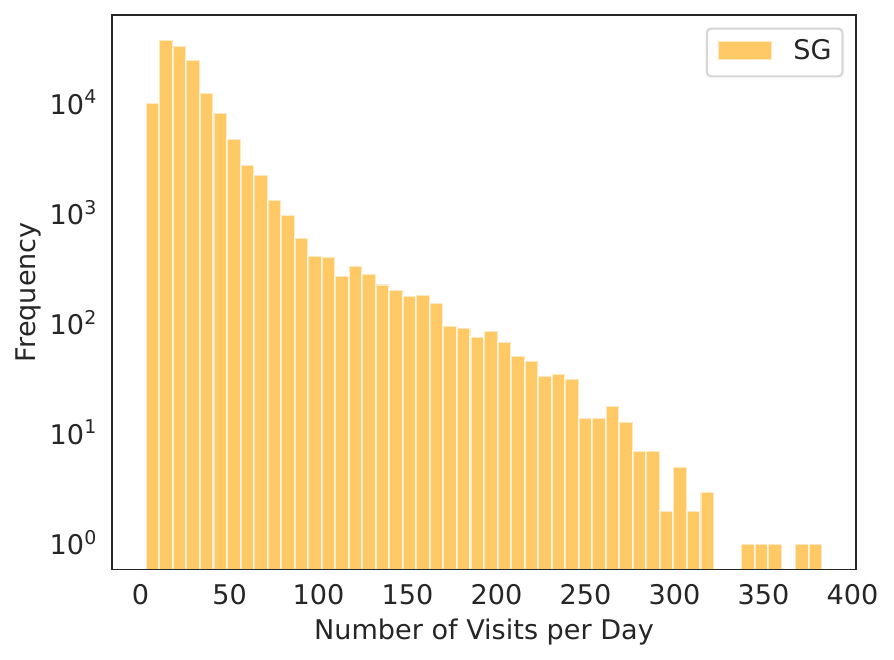}}
    \caption{The distribution plots of three sub-sets: (a) the City Temperature (CT) sub-set includes negative values; (b) the Electricity Consumption Load (ECL) sub-set covers large numbers spanning a wide range; and (c) the sub-set of SafeGraph Human Mobility Data (SG) involves relatively small values ranging from 0 to 400.}
    \label{fig:dis}
\end{figure*}

To highlight the diversity of \name, we analyze its key statistics as given in Table~\ref{tab:dataset}.
\name\ dataset comprises a total of 311,932 instances from three different forecasting application domains.
Each sub-set has its distinct statistical characteristics. 
The last row of Table~\ref{tab:dataset} lists the value range distributions and the average value of each sub-set.
% It is noticeable that CT involves negative values in the dataset and ECL includes large numbers with a wide range.
Furthermore, Figure~\ref{fig:dis} shows the distribution plots of three sub-sets in our \name\ dataset.
It can be observed that the distributions of the three sub-sets are different. 
As shown in the figure and according to the value ranges reported in Table~\ref{tab:dataset}, these three selected data sources ensure the data diversity of our \name\ datasets.
Specifically, the proposed \name\ covers a range of negative numerical values (the CT sub-set), large values (the ECL sub-set), and small/regular values (the SG sub-set).
This diversity of data in our dataset ensures the representativeness of our \name\ dataset and provides a comprehensive benchmark for the performance of different models under different scenarios.

\begin{table}[]
\centering
\caption{\name\ dataset overview and key statistics.}
\scriptsize
\label{tab:dataset}
\addtolength{\tabcolsep}{-1.05ex}
\begin{tabular}{l||c|c|c} \toprule
 & CT & ECL & SG \\ \midrule
Objects-of-interest & 110 cities & 50 Users & 324 POIs \\ \midrule
\multirow{2}{*}{Collection Period} & From 2017/01/01 & From 2012/01/01 & From 2020/06/15 \\
 & To 2020/04/30 & To 2014/12/31 & To 2021/09/05 \\ \midrule
\multirow{4}{*}{Training Set} & From 2017/01/01 & From 2012/01/01 & From 2020/06/15 \\
 & To 2019/04/30 & To 2017/01/31 & To 2021/04/23 \\
 & 850 days & 762 days & 313 days \\
 & 91850 instances & 37350 instances & 96552 instances \\ \midrule
\multirow{4}{*}{Validation Set} & From 2019/05/01 & From 2014/02/01 & From 2021/04/24 \\
 & To 2019/08/31 & To 2014/05/31 & To 2021/06/07 \\
 & 123 days & 120 days & 45 days \\
 & 11880 instances & 5250 instances & 9720 instances \\ \midrule
\multirow{4}{*}{Test Set} & From 2019/09/01 & From 2014/06/01 & From 2021/06/08 \\
 & To 2020/04/30 & To 2014/12/31 & To 2021/09/05 \\
 & 243 days & 214 days & 90 days \\
 & 25080 instances & 9950 instances & 24300 instances \\ \midrule
Value Range & {[}-44, 104{]} & {[}2799, 24906{]} & {[}3, 383{]} \\
Average Value & 58.070 & 11479.120 & 29.355 \\ \bottomrule
\end{tabular}
\end{table}

\section{Benchmark}
In this section, we present our benchmarking study and analysis for the proposed \tname\ task. By conducting experiments on the established \name\ dataset, we aim to address the following two main research questions:
% \begin{itemize}
    % \item 
    \textbf{\textit{RQ1}}: Can we use language generation models to predict time series? Compared to the conventional numerical-based time-series forecasting methods, what is the performance of our PromptCast?
    % \item 
    \textbf{\textit{RQ2}}: Can forecasting time series with prompts as well as using language generation models achieve better generalization ability?
% \end{itemize}

\subsection{Evaluation Metrics}
While the proposed \tname\ task is a language generation task that aims to generate the target output prompts, we are particularly interested in evaluating the time series forecasting performance. 
To this end, the first step of the evaluation protocol is to extract the predicted numerical values from the generated sentences. Given that the output prompts follow the same template for each sub-set (\eg, ``There will be ...'' in the SG sub-set), the numerical value can be easily extracted by simple string parsing.
However, in practice, due to the uncertainty of the inference process, it is not guaranteed that the numerical value can be extracted from the generated output for every testing instance.
To reflect this in the evaluation, we explicitly introduce the Missing Rate as one evaluation metric. 
It is defined as $(n_{\text{test}}-n_{\text{decoded}})/n_{\text{test}}\times 100\%$ where $n_{\text{test}}$ and $n_{\text{decoded}}$ are the total number of instances in the test set and the number of generated instances that can successfully decode the predicted value, respectively. A lower Missing Rate indicates better performance.

After extracting the predicted numerical values, the evaluation of the \tname\ task will be similar to the evaluation of traditional numerical-based forecasting methods. Therefore, we also use two widely-used metrics, the Root Mean Square Error (RMSE) and the Mean Absolute Error (MAE), to evaluate the prediction performance in our benchmark.
For each deep learning method, we report the average performance and the standard deviation over five runs with different random seeds. The Missing Rate is not considered when evaluating numerical-based methods as it is not applicable.

\subsection{Baselines}

In order to provide a comprehensive benchmark for the proposed \tname\ task, we test the performance of 10 popular natural language generation models on our \name\ dataset (\ie, \name-prompt). These language models are T5~\cite{t5}, Bart~\cite{bart}, BERT~\cite{bert}, RoBERTa~\cite{liu2019roberta}, Electra~\cite{ELECTRA}, Bigbird~\cite{zaheer2020big}, ProphetNet~\cite{ProphetNet}, LED~\cite{beltagy2020longformer}, Blenderbot~\cite{blender}, and Pegasus~\cite{zhang2020pegasus}.
Furthermore, for the comparison purpose (RQ1) and providing strong baselines for forecasting with prompts methods, we also include the performance of conventional numerical paradigm forecasting methods on the \name-numerical.
We consider 3 naive forecasting methods: Copy Yesterday (CY), Historical Average (HA), and Copy Last Week (CLW). 
We also consider 3 basic numerical forecasting methods: AutoARIMA, LSTM, and temporal convolutional network (TCN). 
Transformer-based forecasting methods including the vanilla Transformer~\cite{vaswani2017attention}, the state-of-the-art Informer~\cite{zhou2021informer}, Autoformer~\cite{xu2021autoformer}, and FEDformer~\cite{FEDformer} are also included.
% Overall, we consider 20 different methods (10 numerical methods and 10 language models) in  benchmark.

\begin{table*}[]
\centering
\caption{Results of numerical-based forecasting methods on \name-numerical.}
\label{tab:res_numerical}
\scriptsize
\addtolength{\tabcolsep}{-0.325ex}
\begin{tabular}{l|c|c|c|c|c|c|c} \toprule
\multirow{2}{*}{Method} & Temporal & \multicolumn{2}{c|}{CT} & \multicolumn{2}{c|}{ECL} & \multicolumn{2}{c}{SG} \\ \cline{3-8}
 & Embedding & RMSE & MAE & RMSE & MAE & RMSE & MAE \\ \midrule
CY & N/A & 6.710 & 4.991 & 680.142 & 381.247 & 10.945 & 7.691 \\
HA & N/A & 8.089 & 6.321 & 694.658 & 455.288 & 9.198 & 6.221 \\
CLW & N/A & 10.352 & 7.950 & 835.590 & 553.485 & 10.387 & 7.381 \\ \midrule
AutoARIMA & N/A & 6.904 & 5.234 & 644.253 & 387.608 & 9.290 & 6.383 \\
LSTM & N/A & 6.511$\pm$0.053 & 4.956$\pm$0.056 & 598.962$\pm$2.027 & 367.798$\pm$2.088 & 8.994$\pm$0.032 & 6.107$\pm$0.011 \\
TCN & N/A & 6.397$\pm$0.089 & 4.876$\pm$0.072 & 589.785$\pm$6.280 & 368.682$\pm$6.077 & 8.389$\pm$0.029 & 5.927$\pm$0.039 \\ \midrule
 & timeF & 6.790$\pm$0.072 & 5.238$\pm$0.058 & 612.102$\pm$25.081 & 400.182$\pm$24.956 & 8.230$\pm$0.029 & 5.851$\pm$0.023 \\ 
Transformer & fixed & 6.603$\pm$0.177 & 4.989$\pm$0.137 & 557.813$\pm$22.754 & 357.253$\pm$6.875 & 8.274$\pm$0.035 & 5.856$\pm$0.036 \\ 
 & learned & 6.873$\pm$0.143 & 5.294$\pm$0.108 & 567.307$\pm$10.261 & 394.226$\pm$8.900 & 8.408$\pm$0.274 & 5.940$\pm$0.103 \\ \midrule
 & timeF & 6.778$\pm$0.085 & 5.195$\pm$0.075 & 597.011$\pm$15.373 & 383.704$\pm$21.694 & 8.167$\pm$0.049 & 5.832$\pm$0.032 \\ 
Informer & fixed & 6.457$\pm$0.268 & 4.922$\pm$0.209 & \textbf{536.921$\pm$33.375} & \textbf{349.331$\pm$11.916} & \textbf{8.151$\pm$0.068} & 5.868$\pm$0.049 \\ 
 & learned & 6.844$\pm$0.106 & 5.307$\pm$0.083 & 561.661$\pm$19.709 & 394.813$\pm$13.871 & 8.403$\pm$0.281 & 5.914$\pm$0.133 \\ \midrule
 & timeF & 6.681$\pm$0.094 & 5.040$\pm$0.081 & 608.499$\pm$9.051 & 384.782$\pm$9.361 & 8.180$\pm$0.020 & \textbf{5.831$\pm$0.017} \\ 
Autoformer & fixed & 6.438$\pm$0.064 & 4.909$\pm$0.064 & 588.466$\pm$9.446 & 375.703$\pm$8.107 & 8.239$\pm$0.053 & 5.898$\pm$0.025 \\ 
 & learned & 6.812$\pm$0.091 & 5.200$\pm$0.072 & 593.071$\pm$3.476 & 393.695$\pm$2.385 & 8.392$\pm$0.220 & 6.044$\pm$0.158 \\ \midrule
 & timeF & 6.567$\pm$0.158 & 5.015$\pm$0.130 & 633.060$\pm$7.646 & 401.925$\pm$7.186 & 8.314$\pm$0.081 & 5.941$\pm$0.055 \\ 
FEDformer & fixed & \textbf{6.358$\pm$0.050} & \textbf{4.841$\pm$0.029} & 596.240$\pm$13.169 & 403.764$\pm$12.324 & 8.214$\pm$0.013 & 5.913$\pm$0.024 \\ 
 & learned & 6.650$\pm$0.049 & 5.108$\pm$0.036 & 539.039$\pm$2.878 & 387.422$\pm$1.611 & 8.374$\pm$0.051 & 6.049$\pm$0.049 \\ \bottomrule
\end{tabular}
\end{table*}

\subsection{Implementation Details}

For the evaluated numerical forecasting methods, the implementations are based on the official FEDformer\footnote{\url{https://github.com/MAZiqing/FEDformer}} repository which also includes the implementation of Transformer, Informer, and Autoformer. We follow the default hyperparameter settings except for the \textit{pred\_len} parameter (used in Informer, FEDformer, and Autoformer). 
For numerical methods, the data normalization process is also included when we processed the numerical data.
In our experiments, the \textit{pred\_len} parameter is set to 7 (around half of the observation length, which is 15 in \name).
As for the hyperparameter searching, the factor $c$ (used in Informer and Autoformer) is the one that could affect the forecasting performance of numerical forecasting methods~\cite{xu2021autoformer}). The default factor given by the official implementation is 3 and this default factor has also been used for different datasets according to the official implementations. This justifies that it is a reasonable choice when we evaluate Informer and Autoformer performance on our \name\ dataset. 

For the language models in the benchmark, their implementations can be grouped into two categories.
The first category follows the \textit{EncoderDecoderModel}\footnote{\url{https://huggingface.co/docs/transformers/model\_doc/encoder-decoder}} framework. Three language models are implemented under this category: BERT, RoBERTa, Electra.
The rest 7 language models are accomplished through the second category, which is the \textit{ConditionalGeneration} in HuggingFace (\eg, \textit{BartForConditionalGeneration} class\footnote{\url{https://huggingface.co/docs/transformers/v4.19.2/en/model\_doc/bart\#transformers.BartForConditionalGeneration}}).
The fine-tuning process in \tname\ is based on the standard Trainer provided by HuggingFace. Specifically, the sequence-to-sequence trainer is applied as our downstream time series forecasting task in \tname\ is a sequence-to-sequence task. Additionally, no modifications are introduced to the loss function during fine-tuning.
For decoding the generation from the language models, the standard tokenizers provided by HuggingFace are used to detokenize the direct output tokens to yield sentences and there are no extra regularization steps on the yielded sentences to acquire numerical predicted outputs. We simply decode the numerical values through string parsing.

We would like to emphasize that the language models and numerical forecasting methods are treated and processed equally and fairly in our benchmark. There is also no specific hyperparameter tuning for our\tname. For both language models and numerical models, we all use the default settings provided/recommended by the official implementations. Thus, the comparison is fair and convincing. This hyperparameter searching-free characteristic could also reflect another benefit of \tname\ in real-world applications, that is, no need to conduct complicated and time-consuming hyperparameter tuning processes. The forecasting models can then be deployed more quickly for new forecasting scenarios. 
All the above scripts are provided in our GitHub\footnote{\url{https://github.com/HaoUNSW/PISA}}. The experiments were performed with PyTorch on a Linux server equipped with Nvidia V100 GPUs. In our experiments, only 1 GPU is enabled per run for each method.

\begin{table*}[]
\centering
\caption{Results (RMSE and MAE) of our \tname\ using different language models on \name-prompt.}
\label{tab:res_prompt}
% \scriptsize
% \addtolength{\tabcolsep}{-0.7ex}
\begin{tabular}{l|cc|cc|cc|cc|cc|cc} \toprule
 & \multicolumn{4}{c|}{CT} & \multicolumn{4}{c|}{ECL} & \multicolumn{4}{c}{SG} \\ \midrule
 & \multicolumn{2}{c|}{RMSE} & \multicolumn{2}{c|}{MAE} & \multicolumn{2}{c|}{RMSE} & \multicolumn{2}{c|}{MAE} & \multicolumn{2}{c|}{RMSE} & \multicolumn{2}{c}{MAE} \\
 & mean & std & mean & std & mean & std & mean & std & mean & std & mean & std \\ \midrule
T5 & 6.499 & 0.065 & 4.830 & 0.038 & 527.425 & 10.280 & 353.450 & 2.696 & 8.450 & 0.037 & 5.879 & 0.020 \\
Bart & 6.432 & 0.040 & 4.759 & 0.027 & 527.350 & 10.608 & 355.390 & 2.751 & 8.279 & 0.053 & \textbf{5.785} & 0.023 \\
Blenderbot & 6.667 & 0.048 & 4.828 & 0.025 & 541.713 & 10.838 & 355.846 & 4.154 & 8.429 & 0.080 & 5.798 & 0.022 \\
LED & 6.376 & 0.036 & 4.730 & 0.025 & 540.924 & 16.542 & 367.276 & 6.742 & 8.277 & 0.072 & 5.787 & 0.036 \\
Pegasus & 6.379 & 0.023 & 4.727 & 0.014 & 537.186 & 11.296 & 361.135 & 4.728 & 8.289 & 0.016 & 5.817 & 0.013 \\
ProphetNet & 6.375 & 0.063 & 4.740 & 0.052 & 584.814 & 4.124 & 356.632 & 2.712 & 8.466 & 0.135 & 5.847 & 0.071 \\
Bigbird & \textbf{6.351} & 0.016 & \textbf{4.707} & 0.019 & \textbf{519.665} & 3.440 & \textbf{350.699} & 1.953 & 8.326 & 0.048 & 5.841 & 0.031 \\
Electra & 6.397 & 0.011 & 4.740 & 0.013 & 576.506 & 3.789 & 352.187 & 3.413 & 8.311 & 0.084 & 5.820 & 0.046 \\
BERT & 6.388 & 0.081 & 4.758 & 0.052 & 577.076 & 3.608 & 354.653 & 2.169 & 8.395 & 0.040 & 5.823 & 0.030 \\
RoBERTa & 6.450 & 0.081 & 4.786 & 0.070 & 659.874 & 23.218 & 448.902 & 19.320 & \textbf{8.260} & 0.031 & \textbf{5.785} & 0.009 \\ \hline
Numerical Best & 6.358 & 0.020 & 4.841 & 0.029 & 536.921 & 33.375 & {349.331} & 11.916 & {8.151} & 0.068 & 5.831 & 0.017 \\
\bottomrule
\end{tabular}
\end{table*}
% \vspace{-2ex}

\begin{table}[]
\centering
\caption{The Missing Rate performance of language models.}
\small
\label{tab:res_prompt_miss}
\addtolength{\tabcolsep}{-0.3ex}
\begin{tabular}{l|ccc} \hline
Missing Rate & ProphetNet & Electra & BERT \\ \cline{2-4}
 on CT  (\%) & 0.412  $\pm$ 0.045 & 0.319 $\pm$ 0.068 & 0.244  $\pm$ 0.151 \\ \hline
\end{tabular}
\end{table}

\subsection{Experimental Performance}

\subsubsection{Numerical-Based Methods}
This part focuses on evaluating the typical numerical-based forecasting methods with our \name\ dataset. 
Normally, the position embeddings used in the Transformer architecture (and its variants) only contain the limited position information (\eg, the first time step in each input sequence). This kind of position information remains the same for all different input data instances.
However, for time series data, temporal information (\eg, \textit{day-of-week} and \textit{month-of-year}) is an important cue for predicting future data and reflects the global position relations.
For example, the first time step of instance A could correspond to Monday whereas the first time step (same position) of instance B could be Friday.
Thus, appending the temporal embeddings to the basic position embedding becomes popular in Transformer-based time series forecasting methods.
This is equivalent to providing temporal context in the input prompt (\ie, From \{$t_1$\} to \{$t_{\text{obs}}$\} in Table~\ref{tab:dataset_template}) in \tname.
Based on the implementations of Informer\footnote{https://github.com/zhouhaoyi/Informer2020/blob/main/models/embed.py} and Autoformer\footnote{https://github.com/thuml/Autoformer/blob/main/layers/Embed.py}, we fully investigate and benchmark three different embedding approaches, namely, \textit{timeF}, \textit{fixed}, and \textit{learned}.
For the three different temporal embeddings used in numerical methods benchmarking, we also follow the Autoformer implementations\footnote{\url{https://github.com/thuml/Autoformer/blob/main/layers/Embed.py}}. Basically, the \textit{timeF} embedding is accomplished via \textit{nn.Linear()} function and the \textit{fixed} and \textit{learned} are based on \textit{nn.Embedding()} function. The \textit{fixed} embedding has fixed non-trainable parameters (similar to the original sin/cos position embedding weight calculation in the vanilla Transformer) for the \textit{nn.Embedding()} layer, whereas the parameters for the \textit{learned} embedding are trainable.
In this way, the three different embedding methods can be used for numerical forecasting methods to capture the temporal information in the input data and this offers a comprehensive evaluation of the temporal embeddings for the forecasting task.

Table~\ref{tab:res_numerical} presents the performance of different methods with different temporal embedding policies. In general, FEDformer, Informer and Autoformer achieve the best performance across different sub-sets. In most cases, these advanced time series forecasting frameworks outperform the vanilla Transformer, naive methods, and non-Transformer methods.
Naive methods demonstrate worse forecasting performance compared to other methods, which is as expected.
When comparing different embeddings, the \textit{fixed} embedding demonstrates overall good performance. This embedding leads to good predictions on 5 out of 6 metrics and the \textit{timeF} is the best performer of the remaining metric (MAE on SG). The \textit{learned} embedding has the worst performance on CT and SG, but it beats the \textit{timeF} on the ECL sub-set. These results suggest that the \textit{fixed} embedding is a favorable approach for incorporating temporal cues.

\subsubsection{Pre-trained Language Models}

\begin{table}[]
\centering
\caption{Details of HuggingFace Pre-trained Models}
\label{tab:detail_hf}
\footnotesize
\addtolength{\tabcolsep}{-0.875ex}
\begin{tabular}{c||c|c} \toprule
Model & HuggingFace Key & Pretrained Model Size \\ \midrule
T5 & t5-base & 891.7 MB \\
Bart & facebook/bart-base & 557.8 MB \\
Blenderbot & facebook/blenderbot\_small-90M & 350.4 MB \\
LED & allenai/led-base-16384 & 647.7 MB \\
Pegasus & google/pegasus-xsum & 2.3 GB \\
ProphetNet & microsoft/prophetnet-large-uncased & 1.6 GB \\
Bigbird & google/bigbird-pegasus-large-arxiv & 2.3 GB \\
Electra & google/electra-base-generator & 135.0 MB \\
BERT & bert-base-uncased & 440.5 MB \\
RoBERTa & roberta-base & 501.2 MB \\ \bottomrule
\end{tabular}
\end{table}

For language models investigated in the benchmark, the ready-to-use pre-trained weights provided by HuggingFace~\cite{wolf-etal-2020-transformers} are used for initialization.
The configuration details are listed in Table~\ref{tab:detail_hf} and with the model key given in the table, the corresponding pre-trained model can be accessed and downloaded from HuggingFace.
It is worth noting that the pre-trained weights are trained with general English-language corpora datasets such as BookCorpus~\cite{zhu2015aligning}, CC-News~\cite{liu2019roberta}, and OpenWebText~\cite{radford2019language}. These common language datasets are about general articles and do not include specific time series sequences orientated data. 
Although three data sources in \name\ are open available, only the original numerical data (in the csv format) can be acquired online instead of the text format with our prompts. This also guarantees that the data from these three datasets are not used in pre-training and prevents the potential data leakage. 
In the experiments, each language model is fine-tuned with the training set of each sub-set in \name. 

The prediction results (RMSE and MAE) of using different language generation models on \name\ dataset are listed in Table~\ref{tab:res_prompt}.
According to the table, the top performers (shown in bold) include Bigbird, Bart, and RoBERTa. Bigbird achieves the best performance on 4 out of 6 metrics.
When we jointly consider Table~\ref{tab:res_numerical} and Table~\ref{tab:res_prompt}, it can be seen that using language models performs reasonably well on the CT and ECL sub-sets. 
% Specifically, 7 language models outperform the best numerical performer on CT. 
For ECL, although the MAE of using language models is slightly worse than the best performer in Table~\ref{tab:res_numerical}, the RMSE has a relatively large improvement.
Compared with numerical methods, using language models can also yield comparable results on SG.
This benchmark answers RQ1 and indicates that prompt-based forecasting with language models is a promising direction for time series forecasting research.

\begin{table*}
\centering
\begin{threeparttable}[]
\caption{Results of numerical forecasting methods and PromptCast under different settings.}
\label{tab:zeroshot}
\scriptsize
\addtolength{\tabcolsep}{-0.95ex}
\begin{tabular}{l|c|cc|cc|cc|cc|cc|cc} \toprule
\multirow{3}{*}{Method} & \multirow{3}{*}{\begin{tabular}[c]{@{}c@{}}Temporal\\  Embedding\end{tabular}} & \multicolumn{4}{c|}{CT} & \multicolumn{4}{c|}{ECL} & \multicolumn{4}{c}{SG} \\ \cline{3-14}
 &  & \multicolumn{2}{c|}{RMSE} & \multicolumn{2}{c|}{MAE} & \multicolumn{2}{c|}{RMSE} & \multicolumn{2}{c|}{MAE} & \multicolumn{2}{c|}{RMSE} & \multicolumn{2}{c}{MAE} \\ \cline{3-14}
 &  & mean & std & mean & std & mean & std & mean & std & mean & std & mean & std \\ \midrule
 & timeF & 75.465 & 1.330 & 73.238 & 1.473 & 11866.762 & 40.561 & 11288.860 & 41.504 & 29.010 & 2.554 & 18.903 & 1.087 \\
Transformer & fixed & 67.964 & 12.021 & 65.991 & 12.311 & 5780.931 & 1432.223 & 5055.838 & 1836.453 & 52.461 & 17.611 & 47.150 & 21.680 \\
 & learned & 48.691 & 14.586 & 40.968 & 17.008 & 7938.621 & 550.239 & 6982.758 & 647.932 & 28.238 & 1.348 & 18.719 & 1.743 \\ \midrule
 & timeF & 67.783 & 15.014 & 64.901 & 16.422 & 11887.368 & 30.596 & 11306.690 & 32.765 & 34.927 & 3.421 & 25.205 & 3.983 \\
Informer & fixed & 69.109 & 8.656 & 67.065 & 9.090 & 11180.022 & 296.532 & 10649.465 & 259.677 & 26.761 & 2.290 & 15.930 & 1.857 \\
 & learned & 45.517 & 17.482 & 38.000 & 17.228 & 11509.084 & 113.513 & 10923.215 & 114.072 & 27.417 & 2.241 & 17.310 & 1.471 \\ \midrule
 & timeF & 52.814 & 5.002 & 39.577 & 5.842 & 694.693 & 2.715 & 455.658 & 2.188 & 38.710 & 11.207 & 30.857 & 9.751 \\
Autoformer & fixed & 47.691 & 5.329 & 34.531 & 2.996 & 674.641 & 1.845 & 440.564 & 1.678 & 36.801 & 3.523 & 28.637 & 1.927 \\
 & learned & 83.349 & 9.332 & 59.951 & 7.855 & 693.810 & 0.719 & 454.691 & 0.644 & 56.787 & 3.050 & 40.890 & 2.004 \\ \midrule
 & timeF & 63.851 & 4.729 & 46.117 & 4.608 & 693.017 & 2.127 & 454.284 & 1.983 & 50.252 & 8.780 & 40.091 & 8.115 \\
FEDformer & fixed & 77.699 & 3.711 & 54.176 & 4.005 & 655.196 & 3.142 & 424.823 & 2.603 & 64.622 & 5.056 & 45.391 & 2.996 \\
 & learned & 239.426 & 24.961 & 146.535 & 21.858 & 694.019 & 0.832 & 454.866 & 0.842 & 108.169 & 8.851 & 85.243 & 6.055 \\ \midrule \midrule
\multicolumn{14}{c}{Train-from-Scratch} \\ \midrule
PromptCast (Bart) &  & 6.886 & 0.052 & 5.130 & 0.039 & \textbf{546.881} & 8.904 & \textbf{371.102} & 4.369 & \textbf{8.923} & 0.466 & \textbf{6.000} & 0.174 \\
PromptCast (Pegasus) & N/A  & 6.791\tnote{a} & 0.190 & 5.043 & 0.114 & 632.351 & 14.745 & 411.915 & 6.316 & 9.484 & 0.372 & 6.180 & 0.078 \\
PromptCast (Bigbird) &  & \textbf{6.643} & 0.076 & \textbf{4.964} & 0.085 & 639.889 & 8.340 & 416.529 & 4.021 & 10.529 & 0.437 & 6.365 & 0.031 \\ \midrule \midrule
\multicolumn{14}{c}{Zero-Shot} \\ \midrule
PromptCast (Bart) &  & 7.379 & 0.086 & 5.501 & 0.067 & 660.082 & 16.205 & 493.035 & 18.166 & \textbf{8.592} & 0.075 & \textbf{5.961} & 0.038 \\
PromptCast (Pegasus) & N/A  & \textbf{6.918} & 0.022 & \textbf{5.178} & 0.031 & \textbf{643.483} & 16.536 & 446.876 & 5.822 & 9.293 & 0.160 & 6.116 & 0.041 \\
PromptCast (Bigbird) &  & 7.070 & 0.074 & 5.248 & 0.044 & 665.191 & 55.176 & \textbf{417.634} & 4.815 & 9.439 & 0.020 & 6.289 & 0.027 \\ \bottomrule
\end{tabular}
\begin{tablenotes}
  \item[a] Pegasus Missing Rate (\%) on CT: 2.482$\pm$3.754
  \end{tablenotes}
\end{threeparttable}
\end{table*}

Table~\ref{tab:res_prompt_miss} reports the Missing Rate metric. It is clearly noticeable that only three methods (ProphetNet, Electra, and BERT) have a tiny amount (less than 0.5\%) of missing cases and all cases appear on the CT sub-set. 
For other methods that are not listed in Table~\ref{tab:res_prompt_miss}, the missing rates are all 0.
We further investigate the output sentences that cannot be decoded and find out that the failure cases are related and potentially caused by negative values. For example, the failure generated sentences are like \textit{the temperature will be - - - -} where the models fail to generate the tokens after ``-''.
Our \name\ dataset is valuable in supporting the research directions to address this limitation in the future.
For more qualitative examples, we provide all generated examples of language models in our GitHub repository\footnote{\url{https://github.com/HaoUNSW/PISA/PromptCast\%20Generated\%20Examples}} including all the three test sets with various settings such as with pre-trained weights, train-from-scratch, zero-shot, prompt ablations, and multi-step forecasting.

\subsubsection{Performance of ChatGPT}

To showcase the time series forecasting capabilities of recent GPT models, we employed the OpenAI API to assess the performance of GPT-3.5 (``gpt-3.5-turbo'', accessed in August 2023) using identical PISA sub-sets.
Given the prohibitively high expenses associated with using GPT-4 and fine-tuning GPT-3.5 services offered by the OpenAI API, this section of our experiment focused on GPT-3.5 without fine-tuning.
During the evaluation process, we employed the input prompts from the testing sets in PISA as queries and collected the responses from the API. The resulting performance of GPT-3.5 is summarized in Table~\ref{tab:gpt}. When compared to the results reported in the main paper for numerical forecasting methods and our PromptCast, GPT-3.5 exhibits higher RMSE and MAE values across all three sub-sets.
From the table, we can also clearly notice that the GPT-3.5 has large Missing Rates, especially on the SG sub-set. 
We observed numerous outputs containing text such as \textit{``it is difficult to predict the consumption''} for the ECL sub-set and \textit{``we cannot determine the number of people''} for the SG sub-set.
In most cases that GPT-3.5 can make predictions, the predicted values are simply the averaging of the input values of the observation period.
The complete set of outputs for the three sub-sets can be found in our GitHub repository for further detailed inspection. 
\begin{table}[]
\centering
\caption{Results of GPT-3.5 through the OpenAI API.}
\label{tab:gpt}
\begin{tabular}{l|ccc} \hline
    & RMSE     & MAE      & MissingRate \\ \hline
CT  & 10.349   & 6.424    & 0.889\%     \\
ECL & 9413.488 & 1321.210 & 9.879\%     \\
SG  & 21.142   & 6.827    & 82.819\%   \\ \hline
\end{tabular}
\end{table}

\subsubsection{Computation Cost}

\begin{table}[]
\centering
\caption{The comparison of deployment cost in USD and inference speed.}
\label{tab:cost}
\tiny
\addtolength{\tabcolsep}{-0.85ex}
\begin{tabular}{l|c|cc|cc|cc} \toprule
                   & \multirow{2}{*}{\# params} & \multicolumn{2}{c|}{CT} & \multicolumn{2}{c|}{ECL} & \multicolumn{2}{c}{SG}              \\
                   &                            & time   & cost   & time   & cost     & time   & cost   \\ 
                   &                            & seconds   & $\times 10^{-5}$  & seconds   & $\times 10^{-5}$   & seconds & $\times 10^{-5}$ \\ \midrule
Transformer        & 10.6 M                     & 0.058     & 4.902      & 0.062     & 5.303       & 0.055   & 4.704     \\
Informer           & 11.4 M                     & 0.122     & 10.385      & 0.102     & 8.668       & 0.096   & 8.134     \\
Autoformer         & 10.6 M                     & 0.095     & 8.077      & 0.070     & 5.976       & 0.063   & 5.391     \\
FEDformer          & 11.0 M                     & 0.048     & 4.113      & 0.039     & 3.324       & 0.087   & 7.401     \\ \midrule
PromptCast (Bart)               & 139.4 M                    & 0.044     & 3.757      & 0.036     & 3.033       & 0.088   & 7.468     \\
PromptCast (Bigbird)            & 576.9 M                    & 0.065     & 5.489      & 0.052     & 4.393       & 0.130   & 11.046    \\
PromptCast (Pegasus)            & 568.7 M                    & 0.064     & 5.415      & 0.051     & 4.337       & 0.127   & 10.785    \\ \midrule
GPT3.5 & 175 B                      & 5.811     & 54.047     & 5.872     & 59.311      & 3.514   & 33.334   \\ \bottomrule
\end{tabular}
\end{table}
To further enrich the manuscript, we have compared the number of parameters, the inference execution time (per instance, in seconds), and the deployment cost (per instance, measured in $\times 10^{-5}$ USD) of the Transformer-based numerical forecasting methods, our PromptCast with different language models, and the GPT-3.5 (the model for ChatGPT).
We assessed GPT-3.5's costing performance utilizing the OpenAI API (accessed in August 2023) with the "gpt-3.5-turbo" model. For the remaining methods, we estimated deployment costs based on the utilization of Nvidia V100 GPUs on AWS (\ie, p3.2xlarge instance with one Nvidia V100 is 3.06 USD per hour\footnote{\url{https://aws.amazon.com/ec2/instance-types/p3/?nc1=h_ls}}). The costs of these models are associated with the computation time.
For GPT-3.5, it is costed based on the number of input and output tokens, with run time including API request and response durations.

Table~\ref{tab:cost} provides a comprehensive comparison of various models on the three sub-sets in PISA.
As demonstrated in the table, noticeably, GPT-3.5 (OpenAI API) stands out as an exceptionally large and resource-intensive model, with significantly higher computational requirements and associated costs compared to other models, making it less cost-effective. For our proposed PromptCast, although the language models featuring more parameters, there is no substantial increase in computation times and associated costs when contrasted with Transformer-based numerical forecasting models.
This computation cost comparison further shows PromptCast is not only effective on the forecasting accuracy but also cost-effective.

\subsection{Training From Scratch}\label{sec:scratch}

% \noindent\textbf{Train-from-Scratch.}
% To further explore the language models for prompt-based forecasting, 
Another question we aim to explore is whether the language models in \tname\ can still be used for time series forecasting without the use of pre-trained weights.
Hence, we disable the pre-trained weights and train language models from scratch with the training set of \name\ dataset. 
We select three language models that exhibit excellent forecasting performance on all three subsets (according to Table~\ref{tab:res_prompt}) for examination in this part of the experiment: Bart, Pegasus, and Bigbird. The results are reported in  Table~\ref{tab:zeroshot}.
Comparing the results of using pre-training weights (Table~\ref{tab:res_prompt}), we note that there is a performance decrease for each method when pre-trained weights are not loaded, suggesting that utilizing pre-trained weights could result in further performance improvement.
However, even training from scratch, the language generation models (especially Bart) still can yield comparable prediction results compared to numerical-based methods (Table~\ref{tab:res_numerical}). 
This demonstrates that our proposed \tname\ is robust and not entirely reliant on pre-trained weights.
While training a model from scratch does yield acceptable predictions, leveraging the pre-trained weights of language models can significantly enhance forecasting performance.
From the comparison, we can also notice that the influence of without using pre-trained weights has less impact on Bart, whereas the performance decrease in the case of Pegasus and Bigbird is more substantial. This disparity can be attributed to the marked difference in model size between Bart, Pegasus, and Bigbird, as evidenced in Table VI. Bart's significantly smaller model size indicates a lower number of parameters that require training. Consequently, under the train-from-scratch setting (\ie, without using pre-trained weights), training models with a larger number of parameters becomes relatively challenging, especially when considering the same volume of training data.

% \noindent\textbf{Zero-shot Performance.}
\subsection{Zero-shot Performance}
To further explore the language models for prompt-based forecasting, we conduct an experiment under the zero-shot setting (see Table~\ref{tab:zeroshot}). Specifically, we fine-tune each method on two sub-sets and test the fine-tuned model on the test set of the left sub-set (\eg, fine-tune with the training sets of CT and ECL, test on the test set of SG).
% Note that the pre-trained weights are used for initialization before the fine-tuning.
For the comparison purpose, we also evaluate the Transformer-based numerical methods under the same zero-shot setting and the results are given in Table~\ref{tab:zeroshot} (the upper part).
Similar to the results given in Table~\ref{tab:res_numerical} (the normal training setting), the \textit{fixed} embedding has better performance (5 out of 6 metrics) than the other two embeddings under this challenging zero-shot setting.
Since the date range of the three sub-sets are dissimilar (Table~\ref{tab:dataset}), learnable \textit{learned} and \textit{timeF} temporal embeddings would lead to inferior performance when transferred to unseen scenarios as the learned embeddings may not be appropriate for the new scenarios.

Except for the Autoformer and FEDformer on ECL, the numerical-based methods fail to produce satisfactory predictions under the zero-shot setting. Given the distinct characteristics of the three subsets, such poor performance is as expected for numerical methods. However, for the language models, although the performance is lower than the standard setting and the train-from-scratch setting, prompt-based forecasting can still generate reasonable predictions. This demonstrates a strong generalization ability when time series forecasting is addressed with prompts (RQ2).
This strong zero-shot ability could bring advantages in real-world forecasting applications such as rapid deployment for new forecasting scenarios and cold-start forecasting for scenarios without any historical data. In the future, prompt-based forecasting could also enable the exploration of more complex forecasting scenarios such as predicting energy consumption based on weather temperature or forecasting customer traffic with temperature trends.

\begin{table*}[]
\centering
\caption{Results of our PromptCast with different prompt types.}
\label{tab:res_ablation}
\footnotesize
% \addtolength{\tabcolsep}{-0.85ex}
\begin{tabular}{l|c|cc|cc|cc|cc|cc|cc} \toprule
 &  & \multicolumn{4}{c|}{CT} & \multicolumn{4}{c|}{ECL} & \multicolumn{4}{c}{SG} \\ \cline{3-14}
Prompt & Model & \multicolumn{2}{c|}{RMSE} & \multicolumn{2}{c|}{MAE} & \multicolumn{2}{c|}{RMSE} & \multicolumn{2}{c|}{MAE} & \multicolumn{2}{c}{RMSE} & \multicolumn{2}{c}{MAE} \\ \cline{3-14}
 &  & mean & std & mean & std & mean & std & mean & std & mean & std & mean & std \\ \midrule
 & PromptCast (Bart) & 6.512 & 0.016 & 4.790 & 0.012 & 603.571 & 2.461 & 380.583 & 1.695 & 8.677 & 0.055 & 5.912 & 0.013 \\
Basic & PromptCast (Pegasus) & 6.496 & 0.012 & 4.767 & 0.007 & \textbf{595.942} & 2.246 & \textbf{366.677} & 1.373 & \textbf{8.530} & 0.032 & \textbf{5.879} & 0.010 \\
 & PromptCast (Bigbird) & \textbf{6.478} & 0.033 & \textbf{4.752} & 0.019 & 609.315 & 3.071 & 378.100 & 2.112 & 8.576 & 0.035 & 5.922 & 0.009 \\ \midrule
 & PromptCast (Bart) & 6.564 & 0.082 & 4.830 & 0.083 & 612.487 & 3.660 & 382.931 & 2.509 & 8.686 & 0.117 & 5.937 & 0.021 \\
Minimum & PromptCast (Pegasus) & 6.560 & 0.049 & 4.818 & 0.027 & \textbf{597.459} & 3.763 & \textbf{368.174} & 1.178 & \textbf{8.632} & 0.046 & \textbf{5.915} & 0.016 \\
 & PromptCast (Bigbird) & \textbf{6.496} & 0.042 & \textbf{4.766} & 0.026 & 618.907 & 4.422 & 384.327 & 0.861 & 8.774 & 0.065 & 5.950 & 0.015 \\ \bottomrule
\end{tabular}
\end{table*}

\begin{table}[]
\centering
\caption{Basic prompting templates.}
\label{tab:basic_template}
\footnotesize
\addtolength{\tabcolsep}{-1ex}
\begin{tabular}{c||c|p{2.85in}} \toprule
 &  & Template \\ \midrule
 & \multirow{2}{*}{Input} & The average temperature of was \{$x^m_{t_1: t_{\text{obs}}}$\} degree on each day. \\
\multirow{2}{*}{\rotatebox[origin=c]{90}{CT}} &  & What is the temperature going to be on tomorrow? \\ \cline{2-3}
 & Output & The temperature will be \{$x^{m}_{t_{{\text{obs}}+1}}$\} degree. \\ \midrule
 & \multirow{2}{*}{Input} & The client consumed \{$x^m_{t_1: t_{\text{obs}}}$\} kWh of electricity on each day. \\
\multirow{2}{*}{\rotatebox[origin=c]{90}{ECL}} &  & What is the consumption going to be on tomorrow? \\  \cline{2-3}
 & Output & This client will consume \{$x^{m}_{t_{{\text{obs}}+1}}$\} kWh of electricity. \\ \midrule
 & \multirow{2}{*}{Input} & There were \{$x^m_{t_1: t_{\text{obs}}}$\} people visiting the POI on each day. \\
\multirow{2}{*}{\rotatebox[origin=c]{90}{SG}} &  & How many people will visit the POI on tomorrow? \\  \cline{2-3}
 & Output & There will be \{$x^{m}_{t_{{\text{obs}}+1}}$\} visitors. \\ \bottomrule
\end{tabular}
\end{table}

\subsection{Prompts Ablation Study}\label{sec:ablation}

To further investigate the prompts in the \tname\ task, we conduct an ablation study on our prompting templates and two simplified prompts are developed: (1) \textbf{Basic Prompt}: as shown in Table~\ref{tab:basic_template}, we remove auxiliary information (\eg, date information) and only keep the core information (\ie, the historical observation values $x^m_{t_1: t_{\text{obs}}}$) in the input prompts. The output prompts remain the same.
(2) \textbf{Minimum Prompt}: This is the simplest and the most straightforward version of prompts. We use commas to convert the sequential numerical values $x^m_{t_1: t_{\text{obs}}}$ into comma-delimited strings as the input prompts (\eg, \textit{ ``78, 81, 83, 84, 84, 82, 83, 78, 77, 77, 74, 77, 78, 73, 76''}). For the output prompts, the prediction targets $x^{m}_{t_{{\text{obs}}+1}}$ are directly used as single-word strings (\eg, {\fontfamily{qcr}\selectfont ``78''}).
% \end{itemize}

The prediction results of using these two types of prompts on our \name\ dataset are presented in Table~\ref{tab:res_ablation}. We can notice that both two simplified prompts lead to worse performance compared to the default template reported in the main paper (Table~\ref{tab:dataset_template}). 
While the minimum prompts can produce acceptable outcomes, the basic prompts have been found to have superior performance compared to the minimum prompts with the same language model.
These comparison results demonstrate that: (1) including proper contexts (even if these contexts introduce no extra data, \eg, the basic prompt) in the prompt is beneficial; and (2) the auxiliary information is a significant component in the prompt for better prediction performance.

\subsection{Multi-step Forecasting}\label{sec:multi}

The proposed language foundation models-based \tname\ forecasting paradigm is also suitable for multi-step forecasting scenarios.
% with varying observation lengths (\ie, input lengths) and prediction horizons (\ie, output lengths).
For the multi-step forecasting (\ie, larger prediction horizon $n$), the question part of the input prompt template can be updated to “how many people will visit POI in the next Y days?” and the output prompt is also needed to be revised to reflect the multi-step future data values such as “For day \textit{X} to day \textit{Y}, there will be \textit{A, B, C,} … visitors.”
Here is an example of predicting the next 7-time steps ($n=7$) with the SG sub-set: Input Prompt “\textit{From April 24, 2021, Saturday to May 08, 2021, Saturday, there were 6, 8, 6, 15, 12, 4, 13, 7, 8, 12, 16, 9, 11, 18, 10 people visiting POI 1 on each day. How many people will visit POI 1 in the next 7 days?}” and Output Prompt “\textit{From May 09, 2021, Sunday to May 15, 2021, Saturday, there will be 11, 11, 8, 13, 10, 18, 19 visitors.}” 
% Based on these modified prompts, we also examine the three language models (Bart, Bigbird, Pegasus) with  prediction horizons (4 days, and 7 days).
Building on these modified prompts, we further examine the performance of our PromptCast with three language models (Bart, Bigbird, Pegasus) when using varying observation lengths (input lengths) and prediction horizons (output lengths).
Specifically, we test the models using observation length settings of 7 days and 15 days, and prediction horizons of 1 day, 4 days, and 7 days, resulting in a total of six length configuration combinations.
To handle the multitude of experiments involving different length setting combinations, we exclusively present the results for Transformer-based numerical forecasting methods using the ``fixed" temporal embedding.
The performance of these three language models on these multi-step forecasting settings are listed in Table~\ref{tab:res_multistep}.

From the table, we can see that \tname\ can be easily adapted to multi-step forecasting while using the exact same forecasting model architectures (\ie, the language foundation models) for different settings.
Generally, compared to the single-step forecasting performance, increasing the prediction horizon leads to larger missing rates. 
% Specifically, Bart shows a worse performance as evidenced by the large missing rates on different sub-sets. However, Bigbird and Pegasus can still yield reasonably good predictions with extremely small missing rates, which demonstrates language models are still powerful in predicting multiple steps ahead. 
% The results further confirm that our proposed \tname\ is suitable for different length settings.
Compared to Transformer-based methods, language models exhibit less favorable predictions in the context of "observing 7 days to predict 7 days" and "observing 15 days to predict 7 days," which are notably challenging scenarios. Bart, in particular, displays suboptimal performance, as evidenced by substantial missing rates across various subsets. However, both Bigbird and Pegasus manage to deliver reasonably accurate forecasts with extremely low missing rates in these difficult settings. Especially, the performance of Pegasus on these two settings are quite close to the state-of-the-art Transformer-based methods. 
Additionally, we notice that using a 15-day observation period often leads to better performance when predicting the same future length. This is likely because a 15-day period encompasses weekly patterns in the observation, leading to more accurate forecasting.

% The results further confirm the suitability of our proposed \tname\ model for different length settings. We can observe that, as the prediction horizon increases, the missing rate becomes larger. Despite this, Bigbird and Pegasus still produce relatively accurate predictions with very low missing rates, demonstrating the power of language models in predicting multiple steps ahead. 

To provide further insight into the performance of the multi-step forecasting, additional examples for the next 7 days setting are available in our repository\footnote{\url{https://github.com/HaoUNSW/PISA/Dataset/PISA\_Plus\_Multistep\_examples/README.md}}.
Our exploration of \tname\ on different lengths illustrates that the model is robust to dynamic observation/prediction lengths, and presents a promising research direction for time series forecasting.

% From the table, it is evident that \tname\ can be easily adapted to multi-step forecasting while maintaining the same forecasting model architectures (\ie, language foundation models) across different settings.
% Generally, as the prediction horizon $n$ increases, the missing rate also increases, which is as expected. Specifically, Bart demonstrates a poorer performance, as evidenced by the high missing rates across different sub-sets. However, Bigbird and Pegasus still exhibit reasonably good predictions with minimal missing rates, demonstrating that language models are still effective in predicting multiple steps ahead.

\begin{table*}[]
\centering
\caption{Results of applying different length settings.}
\label{tab:res_multistep}
\tiny
\addtolength{\tabcolsep}{-1.7ex}
\begin{tabular}{ll|l|ccc|ccc|ccc} \toprule
 &  &  &  & CT &  &  & ECL &  &  & SG &  \\
Obs & Pred &  & RMSE & MAE & MissingRate & RMSE & MAE & MissingRate & RMSE & MAE & MissingRate \\ \midrule
 &  & Transformer & 6.717$\pm$0.094 & 5.144$\pm$0.066 & n/a & 545.909$\pm$18.380 & 358.854$\pm$3.179 & n/a & 8.270$\pm$0.021 & 5.906$\pm$0.009 & n/a \\
 &  & Informer & 6.779$\pm$0.082 & 5.171$\pm$0.073 & n/a & 540.269$\pm$31.190 & 358.000$\pm$14.493 & n/a & 8.211$\pm$0.122 & 5.884$\pm$0.039 & n/a \\
 &  & Autoformer & 7.152$\pm$0.099 & 5.472$\pm$0.067 & n/a & 669.884$\pm$18.113 & 467.813$\pm$19.735 & n/a & 9.594$\pm$0.466 & 7.087$\pm$0.412 & n/a \\
7 Days & 1 Day & FEDformer & 6.378$\pm$0.019 & 4.825$\pm$0.018 & n/a & 669.981$\pm$15.411 & 442.723$\pm$16.427 & n/a & 8.677$\pm$0.059 & 6.228$\pm$0.051 & n/a \\
 &  & PromptCast (Bart) & 6.762$\pm$0.090 & 5.017$\pm$0.083 & 0 & 527.772$\pm$7.282 & 360.066$\pm$3.688 & 0 & 8.427$\pm$0.016 & 5.909$\pm$0.023 & 0 \\
 &  & PromptCast (BigBird) & 6.425$\pm$0.024 & 4.716$\pm$0.013 & 0 & 530.929$\pm$4.792 & 365.122$\pm$2.184 & 0 & 8.526$\pm$0.052 & 5.962$\pm$0.026 & 0 \\
 &  & PromptCast (Pegasus) & 6.820$\pm$0.044 & 5.088$\pm$0.031 & 0 & 527.735$\pm$5.921 & 360.702$\pm$2.352 & 0 & 8.406$\pm$0.063 & 5.957$\pm$0.028 & 0 \\ \midrule
 &  & Transformer & 9.084$\pm$0.171 & 7.043$\pm$0.132 & n/a & 722.580$\pm$19.818 & 472.678$\pm$14.923 & n/a & 8.666$\pm$0.029 & 6.117$\pm$0.024 & n/a \\
 &  & Informer & 9.105$\pm$0.078 & 7.059$\pm$0.057 & n/a & 684.407$\pm$30.136 & 459.958$\pm$11.455 & n/a & 8.483$\pm$0.019 & 6.021$\pm$0.016 & n/a \\
 &  & Autoformer & 9.385$\pm$0.545 & 7.404$\pm$0.479 & n/a & 804.546$\pm$22.951 & 563.633$\pm$19.278 & n/a & 10.487$\pm$0.733 & 7.711$\pm$0.654 & n/a \\
7 Days & 4 Days & FEDformer & 8.320$\pm$0.039 & 6.461$\pm$0.049 & n/a & 701.663$\pm$5.688 & 474.881$\pm$5.411 & n/a & 8.900$\pm$0.022 & 6.244$\pm$0.039 & n/a \\
 &  & PromptCast (Bart) & 10.087$\pm$0.064 & 7.478$\pm$0.058 & 1.622\%$\pm$0.721\% & 1075.721$\pm$218.227 & 542.006$\pm$16.379 & 6.791\%$\pm$2.949\% & 10.156$\pm$0.164 & 6.982$\pm$0.101 & 2.655\%$\pm$1.285\% \\
 &  & PromptCast (BigBird) & 9.213$\pm$0.117 & 6.880$\pm$0.073 & 0 & 750.708$\pm$28.990 & 493.466$\pm$9.704 & 0.047\%$\pm$0.015\% & 9.619$\pm$0.824 & 6.654$\pm$0.495 & 0 \\
 &  & PromptCast (Pegasus) & 9.422$\pm$0.080 & 7.003$\pm$0.047 & 0 & 741.931$\pm$9.613 & 482.426$\pm$4.749 & 0.120\%$\pm$0.116\% & 8.813$\pm$0.039 & 6.142$\pm$0.005 & 0 \\ \midrule
 &  & Transformer & 9.577$\pm$0.476 & 7.448$\pm$0.364 & n/a & 760.256$\pm$24.305 & 510.032$\pm$17.605 & n/a & 8.903$\pm$0.058 & 6.235$\pm$0.036 & n/a \\
 &  & Informer & 9.538$\pm$0.326 & 7.423$\pm$0.211 & n/a & 731.748$\pm$34.708 & 198.438$\pm$18.072 & n/a & 8.606$\pm$0.026 & 6.080$\pm$0.030 & n/a \\
 &  & Autoformer & 10.020$\pm$0.279 & 7.852$\pm$0.263 & n/a & 825.750$\pm$12.678 & 580.922$\pm$13.137 & n/a & 9.039$\pm$0.203 & 6.405$\pm$0.168 & n/a \\
7 Days & 7 Days & FEDformer & 8.990$\pm$0.032 & 6.997$\pm$0.025 & n/a & 739.487$\pm$1.591 & 516.934$\pm$2.579 & n/a & 8.959$\pm$0.037 & 6.287$\pm$0.047 & n/a \\
 &  & PromptCast (Bart) & 11.681$\pm$0.183 & 8.620$\pm$0.058 & 22.745\%1.926\% & 1332.287$\pm$266.259 & 612.469$\pm$32.104 & 29.310\%$\pm$20.472\% & 12.453$\pm$0.894 & 8.375$\pm$0.169 & 32.716\%$\pm$5.301\% \\
 &  & PromptCast (BigBird) & 10.200$\pm$0.120 & 7.601$\pm$0.083 & 0.036\%$\pm$0.018\% & 946.462$\pm$57.607 & 551.675$\pm$6.601 & 0.537\%$\pm$0.314\% & 10.851$\pm$0.612 & 7.433$\pm$0.410 & 0.034\%$\pm$0.026\% \\
 &  & PromptCast (Pegasus) & 10.144$\pm$0.188 & 7.526$\pm$0.131 & 0.022\%$\pm$0.014\% & 851.768$\pm$8.422 & 545.536$\pm$2.627 & 3.669\%$\pm$1.086\% & 8.926$\pm$0.017 & 6.247$\pm$0.011 & 0.013\%$\pm$0.005\% \\ \midrule
 &  & Transformer & 6.603$\pm$0.177 & 4.989$\pm$0.137 & n/a & 557.813$\pm$22.754 & 357.253$\pm$6.875 & n/a & 8.274$\pm$0.035 & 5.856$\pm$0.036 & n/a \\
 &  & Informer & 6.457$\pm$0.268 & 4.922$\pm$0.209 & n/a & 536.921$\pm$33.375 & 349.331$\pm$11.916 & n/a & 8.151$\pm$0.068 & 5.868$\pm$0.049 & n/a \\
 &  & Autoformer & 6.438$\pm$0.064 & 4.909$\pm$0.064 & n/a & 588.466$\pm$9.446 & 375.703$\pm$8.107 & n/a & 8.239$\pm$0.053 & 5.898$\pm$0.025 & n/a \\
15 Days & 1 Day & FEDformer & 6.358$\pm$0.050 & 4.841$\pm$0.029 & n/a & 596.240$\pm$13.169 & 403.764$\pm$12.324 & n/a & 8.214$\pm$0.013 & 5.913$\pm$0.024 & n/a \\
 &  & PromptCast (Bart) & 6.432$\pm$0.040 & 4.759$\pm$0.027 & 0 & 527.350$\pm$10.608 & 355.390$\pm$2.751 & 0 & 8.279$\pm$0.053 & 5.785$\pm$0.023 & 0 \\
 &  & PromptCast (BigBird) & 6.351$\pm$0.016 & 4.707$\pm$0.019 & 0 & 519.665$\pm$3.440 & 350.699$\pm$1.953 & 0 & 8.326$\pm$0.048 & 5.841$\pm$0.031 & 0 \\
 &  & PromptCast (Pegasus) & 6.379$\pm$0.023 & 4.727$\pm$0.014 & 0 & 537.186$\pm$11.296 & 361.135$\pm$4.728 & 0 & 8.289$\pm$0.016 & 5.817$\pm$0.013 & 0 \\ \midrule
 &  & Transformer & 9.325$\pm$0.180 & 7.214$\pm$0.154 & n/a & 673.593$\pm$10.473 & 445.729$\pm$5.224 & n/a & 8.946$\pm$0.182 & 6.211$\pm$0.193 & n/a \\
 &  & Informer & 8.951$\pm$0.091 & 6.941$\pm$0.064 & n/a & 657.442$\pm$10.941 & 447.143$\pm$3.776 & n/a & 8.522$\pm$0.117 & 5.989$\pm$0.095 & n/a \\
 &  & Autoformer & 8.937$\pm$0.163 & 6.951$\pm$0.143 & n/a & 745.216$\pm$11.103 & 521.237$\pm$8.988 & n/a & 8.714$\pm$0.131 & 6.119$\pm$0.086 & n/a \\
15 Days & 4 Days & FEDformer & 8.251$\pm$0.062 & 6.431$\pm$0.038 & n/a & 694.728$\pm$4.038 & 478.788$\pm$3.258 & n/a & 8.617$\pm$0.038 & 6.134$\pm$0.053 & n/a \\
 &  & PromptCast (Bart) & 10.135$\pm$0.043 & 7.631$\pm$0.033 & 1.799\%$\pm$1.651\% & 855.906$\pm$60.640 & 528.250$\pm$15.276 & 25.701\%$\pm$3.393\% & 9.476$\pm$0.079 & 6.559$\pm$0.029 & 1.212\%$\pm$0.325\% \\
 &  & PromptCast (BigBird) & 9.432$\pm$0.136 & 7.075$\pm$0.096 & 0 & 700.280$\pm$14.048 & 481.145$\pm$7.026 & 0.041\%$\pm$0.032\% & 9.392$\pm$0.571 & 6.436$\pm$0.313 & 0 \\
 &  & PromptCast (Pegasus) & 9.582$\pm$0.118 & 7.166$\pm$0.085 & 0 & 769.008$\pm$17.279 & 476.013$\pm$3.494 & 0.096\%$\pm$0.112\% & 8.743$\pm$0.062 & 6.082$\pm$0.026 & 0 \\ \midrule
 &  & Transformer & 9.732$\pm$0.114 & 7.542$\pm$0.092 & n/a & 721.983$\pm$6.746 & 491.780$\pm$12.208 & n/a & 9.004$\pm$0.166 & 6.198$\pm$0.082 & n/a \\
 &  & Informer & 9.615$\pm$0.138 & 7.476$\pm$0.099 & n/a & 709.744$\pm$2.346 & 487.892$\pm$3.853 & n/a & 8.528$\pm$0.131 & 5.959$\pm$0.078 & n/a \\
 &  & Autoformer & 9.812$\pm$0.138 & 7.667$\pm$0.077 & n/a & 831.116$\pm$41.423 & 595.765$\pm$45.440 & n/a & 8.726$\pm$0.174 & 6.148$\pm$0.175 & n/a \\
15 Days & 7 Days & FEDformer & 8.746$\pm$0.083 & 6.807$\pm$0.058 & n/a & 732.936$\pm$13.658 & 508.407$\pm$15.869 & n/a & 8.677$\pm$0.079 & 6.158$\pm$0.104 & n/a \\
 &  & PromptCast (Bart) & 12.033$\pm$0.605 & 8.773$\pm$0.109 & 27.107\%$\pm$1.159\% & 1190.564$\pm$111.116 & 611.117$\pm$55.916 & 50.259\%$\pm$28.139\% & 11.048$\pm$0.325 & 7.653$\pm$0.129 & 38.328\%$\pm$5.362\% \\
 &  & PromptCast (BigBird) & 10.329$\pm$0.162 & 7.745$\pm$0.120 & 0.023\%$\pm$0.031\% & 1063.650$\pm$222.601 & 541.538$\pm$7.397 & 0.329\%$\pm$0.190\% & 9.731$\pm$0.427 & 6.658$\pm$0.200 & 0.007\%$\pm$0.011\% \\
 &  & PromptCast (Pegasus) & 10.522$\pm$0.187 & 7.899$\pm$0.142 & 0.179\%$\pm$0.265\% & 835.386$\pm$23.031 & 525.537$\pm$6.067 & 1.422\%$\pm$0.465\% & 8.802$\pm$0.031 & 6.139$\pm$0.016 & 0.011\%$\pm$0.006\%  \\ \bottomrule
\end{tabular}
\end{table*}

\subsection{Multivariate Time Series Forecasting}\label{sec:mts}

\begin{table*}[]
\centering
\caption{Examples of two types of prompts under the multivariate time series setting.}
\label{tab:mts}
\addtolength{\tabcolsep}{-0.25ex}
\begin{tabular}{p{1.25in}|p{4.5in}} \toprule
Input Prompt A\&B & From May 13, 2016, Friday to May 27, 2016, Friday, the PM2.5 was 32, 23, 20, 53, 82, 113, 133, 94, 64, 83, 20, 4, 18, 48, 57; PM10 was 32, 40, 57, 87, 90, 113, 133, 94, 64, 83, 20, 20, 18, 48, 103; and SO2 was 2, 3, 4, 11, 29, 27, 26, 33, 13, 14, 6, 2, 2, 6, 16 on each day. What are the pollutant values going to be on May 28, 2016, Saturday? \\ \midrule
Output Prompt A & The pollutant values will be 3, 23, 2. \\ \midrule
Output Prompt B & The PM2.5 will be 3. The PM10 will be 23. The SO2 will be 2. \\ \midrule
 & From May 13, 2016, Friday to May 27, 2016, Friday, the PM2.5 was 32, 23, 20, 53, 82, 113, 133, 94, 64, 83, 20, 4, 18, 48, 57 on each day. What is the PM2.5 value going to be on May 28, 2016, Saturday?\\
Input Prompt C & From May 13, 2016, Friday to May 27, 2016, Friday, the PM10 was 32, 40, 57, 87, 90, 113, 133, 94, 64, 83, 20, 20, 18, 48, 103 on each day. What is the PM10 value going to be on May 28, 2016, Saturday?\\ 
& From May 13, 2016, Friday to May 27, 2016, Friday, the SO2 was 2, 3, 4, 11, 29, 27, 26, 33, 13, 14, 6, 2, 2, 6, 16 on each day. What is the SO2 value going to be on May 28, 2016, Saturday?\\ \midrule
& The PM2.5 will be 3.\\
Output Prompt C & The PM10 will be 23.\\
& The SO2 will be 2. \\
\bottomrule
\end{tabular}
\end{table*}

% \begin{table}[]
% \centering
% \caption{Results of language models on multivariate time series forecasting.}
% \label{table:res_mts}
% \footnotesize
% \addtolength{\tabcolsep}{-0.5ex}
% \begin{tabular}{l|c||ccc} \toprule
% Prompt &  & RMSE & MAE & MissingRate \\ \midrule
%  & CY & 73.058 & 43.274 & N/A \\
% N/A & HA & 63.289 & 40.772 & N/A \\
%  & CLW & 87.250 & 52.796 & N/A \\ \midrule
%  & Bart & 79.218$\pm$4.948 & 46.553$\pm$2.838 & 0 \\
% Prompt A & BigBird & 75.783$\pm$1.988 & 45.590$\pm$2.105 & 2.681\%$\pm$3.471\% \\
%  & Pegasus & 82.859$\pm$0.588 & 48.168$\pm$0.444 & 0 \\ \midrule
%  & Bart & 75.268$\pm$2.150 & 41.869$\pm$0.915 & 0 \\
% Prompt B & BigBird & 81.510$\pm$2.524 & 48.913$\pm$0.639 & 0 \\
%  & Pegasus & 75.706$\pm$1.736 & 45.206$\pm$1.156 & 0 \\ \midrule
%  & Bart & 68.282$\pm$2.345 & 40.690$\pm$1.131 & 0 \\
% Prompt C & BigBird & 67.729$\pm$1.880 & 41.547$\pm$1.071 & 0 \\
%  & Pegasus & 67.834$\pm$0.859 & 39.893$\pm$0.691 & 0 \\ \bottomrule
% \end{tabular}
% \end{table}

\begin{table}[]
\centering
\caption{Results of language models on multivariate time series forecasting.}
\label{tab:res_mts}
\scriptsize
\addtolength{\tabcolsep}{-0.75ex}
\begin{tabular}{l|c|ccc} \toprule
                  \multicolumn{2}{l|}{ }         & RMSE             & MAE              & MissingRate         \\ \midrule
\multicolumn{2}{l|}{CopyYesterday}        & 73.058           & 43.274           & n/a                 \\
\multicolumn{2}{l|}{HistrocialAverage}         & 63.289           & 40.772           & n/a                 \\
\multicolumn{2}{l|}{CopylastWeek}            & 87.250           & 52.796           & n/a                 \\\midrule \midrule
                  & fixed   & 52.859$\pm$1.127 & 39.341$\pm$1.300 & n/a                 \\
Transformer       & learned & 55.262$\pm$0.850 & 41.147$\pm$1.142 & n/a                 \\
                  & timeF   & 54.839$\pm$1.156 & 40.385$\pm$1.785 & n/a                 \\ \midrule
                  & fixed   & 53.770$\pm$1.295 & 39.952$\pm$1.018 & n/a                 \\
Informer          & learned & 54.881$\pm$0.630 & 40.903$\pm$0.650 & n/a                 \\
                  & timeF   & 54.801$\pm$0.839 & 41.009$\pm$0.890 & n/a                 \\ \midrule
                  & fixed   & 56.693$\pm$0.602 & 42.770$\pm$0.459 & n/a                 \\
Autoformer        & learned & 58.917$\pm$2.084 & 45.011$\pm$1.740 & n/a                 \\
                  & timeF   & 62.864$\pm$2.090 & 48.464$\pm$1.738 & n/a                 \\ \midrule
                  & fixed   & 55.849$\pm$1.291 & 42.470$\pm$1.195 & n/a                 \\
FEDformer         & learned & 55.905$\pm$0.542 & 52.402$\pm$0.588 & n/a                 \\
                  & timeF   & 63.427$\pm$4.349 & 49.619$\pm$4.254 & n/a                 \\ \midrule\midrule
                  & Bart    & 79.218$\pm$4.948 & 46.553$\pm$2.838 & 0                   \\
Prompt A          & BigBird & 75.783$\pm$1.988 & 45.590$\pm$2.105 & 2.681\%$\pm$3.471\% \\
                  & Pegasus & 82.859$\pm$0.588 & 48.168$\pm$0.444 & 0                   \\ \midrule
                  & Bart    & 75.268$\pm$2.150 & 41.869$\pm$0.915 & 0                   \\
Prompt B          & BigBird & 81.510$\pm$2.524 & 48.913$\pm$0.639 & 0                   \\
                  & Pegasus & 75.706$\pm$1.736 & 45.206$\pm$1.156 & 0                   \\ \midrule
                  & Bart    & 68.282$\pm$2.345 & 40.690$\pm$1.131 & 0                   \\
Prompt C          & BigBird & 67.729$\pm$1.880 & 41.547$\pm$1.071 & 0                   \\
                  & Pegasus & 67.834$\pm$0.859 & 39.893$\pm$0.691 & 0   \\ \bottomrule                
\end{tabular}
\end{table}

As the first attempt to leverage language models for time series forecasting, we take the basic forecasting setting to demonstrate the concept of \tname. However, we would like to emphasize that the proposed new paradigm is flexible and suitable for other forecasting settings such as the multivariate time series forecasting setting.
We only need to update the prompt templates accordingly when \tname\ is adapted to multivariate time series forecasting setting.
Under the new setting, although input/output prompts might be updated, the core prediction model (\ie, language foundation models) in \tname can remain the exact same (prompt-agnostic) and same pre-trained weights can also be used. For example, in the experiments under Section~\ref{sec:ablation}, different prompts are applied. But we can also directly use the same models (without any changes to the model structures) to yield predictions. For conventional numerical forecasting methods, however, necessary updates on the deep learning model (\eg, update the encoder part to fit multivariate features) must be introduced when the forecasting setting is changed. Additionally, processes such as extra hyperparameter search are often required when the model structure is modified. This could also reflect the ``code less'' benefits and robustness of the proposed \tname\ paradigm.  

To investigate the capability of the proposed \tname\ on multivariate forecasting, we further conduct a pilot study of multivariate \tname\ on Beijing Air Quality Data\footnote{https://archive.ics.uci.edu/ml/datasets/Beijing+Multi-Site+Air-Quality+Data} in addition to the three sub-sets in the main \name\ dataset.
The data collection period of this Air Quality (AQ) set is from 2013/03/01 to 2017/02/28. Similar to the main \name, we split this period into training set (2013/03/01-2015/12/18), validation set (2015/12/19-2016/05/12), and test set (2016/05/13-2017/02/28).
After filtering missing values, we focus on three types of air pollutants: PM2.5, PM10, and SO2, which means the feature dimension is 3.

To apply \tname\ for multivariate forecasting, we update the prompt template to include multiple features as Table~\ref{tab:res_mts}. 
In total, we design and examine three different prompts.
For Prompt A and B, the input prompt consists of the description of all 3 features and the main difference of these two prompts is in the output prompt part. Output Prompt A jointly describes the three features, whereas the three feature values are given separately in three short sentences in Output Prompt B.
Prompt C can be seen as an extended version of the prompt template (\eg, Table~\ref{tab:dataset_template}) for the univariate forecasting. It decomposes the multivariate features (\ie, three pollutants in this AQ dataset) into several univariate features.

% The input prompt is appended with the description of all 3 features.
% For the output prompt, we specifically investigate two different types of prompts (Output Prompt A and B in the table).
The results of \tname\ with language foundation models using different prompts are reported in Table~\ref{tab:res_mts}.
Additionally, we list the performance of other baselines.
From the table, we can observe that: (1) All three language models can generate plausible predicted future descriptions as evidenced by almost all zero missing rates.
(2) The design of prompts could definitely affect the prediction performance. Specifically, Prompt C leads to better performance than Prompt A and B on this multivariate forecasting dataset. 
(3) PromptCast has a better MAE performance (\eg, Transformer-fixed VS. Prompt C-Pegasus) whereas the performance on RMSE is worse than the Transformer-based baselines.
We think the potential reason of why \tname\ is not the best performer in this comparison is because that the current prompts cannot fully characterize the relation of different features. This also explains why Prompt C outperforms the other two prompts.

\noindent\textbf{Open Question and Future Work.}
The results show that the model architectures used for univariate time series can also be applied to multivariate time series, this is an indication that the \tname\ task is a promising direction for multivariate time series forecasting. However, further research is needed to fully understand the behavior of multivariate time series forecasting with \tname\ and to develop more robust models that can handle this type of data.
The above performance analysis points out an open question for \tname, that is, how to design prompts for the more challenging multivariate time series. 
One potential working direction is to learn the internal correlations between different temporal features and then develop prompts to represent the learned correlations.
In the future, we will focus on this direction and also explore techniques such as learnable prompts for the multivariate \tname.
We hope the proposed \name\ dataset and the corresponding benchmark could encourage other researchers as well to investigate this interesting \tname\ topic.

\section{Case Studies}\label{sec:case}

\textbf{Why Language Model Works for Forecasting.}
Through the above experiments, we have shown that the proposed \tname\ task represents a new paradigm for time-series prediction. In this section, we aim to further investigate the reasons behind why language models can be used for forecasting purposes.
First, the model architecture of language models aligns seamlessly with the nature of time series data, as both time series forecasting and language generation are inherently sequence-to-sequence processes. This alignment provides a solid foundation for employing language models in time series forecasting.
When viewed from the sequence-to-sequence perspective, time series prediction can be seen as similar to the problem of predicting the next sentence in NLP, which has been effectively tackled using pre-trained foundation models.
% This suggests that the model architectures of language foundation models should also be adaptable for time series sequences.
% incorporating contextual information could and using natural language prompts is the most direct way to take semantic context into accounts in the forecasting process.
Second, using auxiliary information such as \textit{time-of-day}, \textit{day-of-week}~\cite{zhou2021informer} and semantic information~\cite{xue2021mobtcast}  has been shown to be beneficial for improving forecasting performance.
However, incorporating such semantic contexts in conventional numerical-based forecasting methods often requires additional efforts to explicitly design special layers or modules in a heuristic way for the auxiliary inputs. 
Even though some approaches (such as the temporal embedding methods discussed above) have been proposed for numerical-based methods to incorporate auxiliary information, it remains challenging for a forecasting model to seamlessly model the contexts and the main temporal information.
In contrast, with \tname, the prompting process effectively integrates contextual information and time series data, transforming the cross-modality relationship between semantic information and numerical values into an in-modality synergy.
% Using natural language prompts, on the other hand, is a more direct way to take semantic information into account in the forecasting process, which presents a new perspective for the forecasting task.
Both auxiliary and core temporal inputs are considered together as tokens after prompting.
The intra-relation of numerical value tokens at different time steps and the inter-relation between numerical value tokens and contextual information tokens (\eg, date information) could be better learned simultaneously by language foundation models (\eg, through the self-attentions in Transformers). Modeling these relations would then result in good forecasting performance.
% \footnote{In Section~\ref{appendix:case}, we conduct case studies to further investigate this.}.

\begin{figure*}
    \centering
    \subfigure[Case 1: CT sub-set.]{
    \includegraphics[width=.85\textwidth]{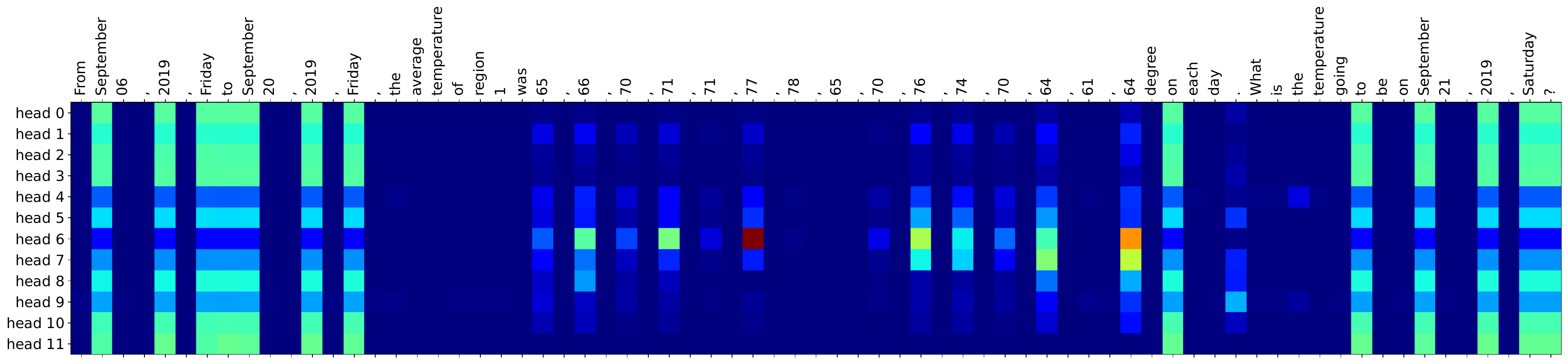}}\\ \vspace{-2ex}
    \subfigure[Case 2 (layer 4): SG sub-set.]{
    \includegraphics[width=.85\textwidth]{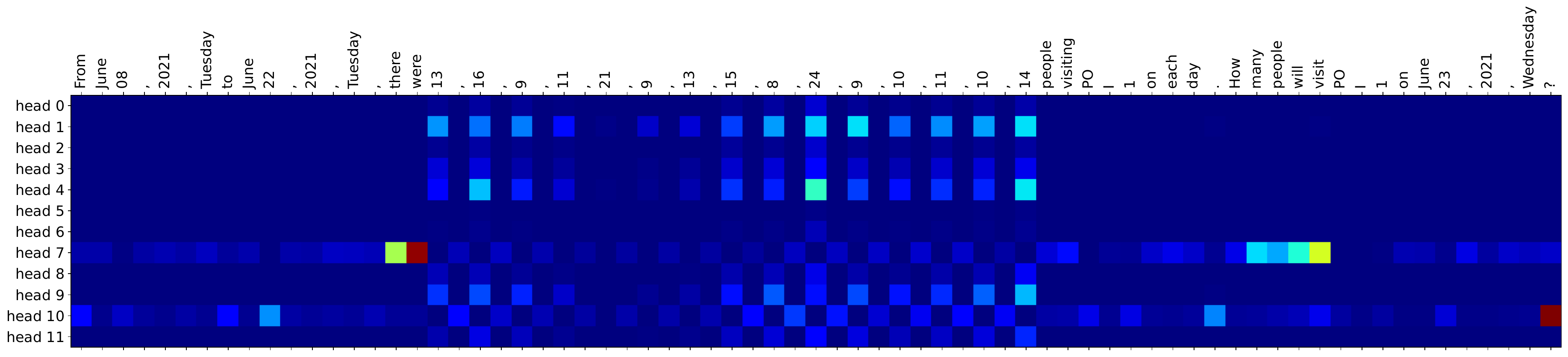}} \\ \vspace{-2ex}
    \subfigure[Case 2 (layer 5): SG sub-set.]{
    \includegraphics[width=.85\textwidth]{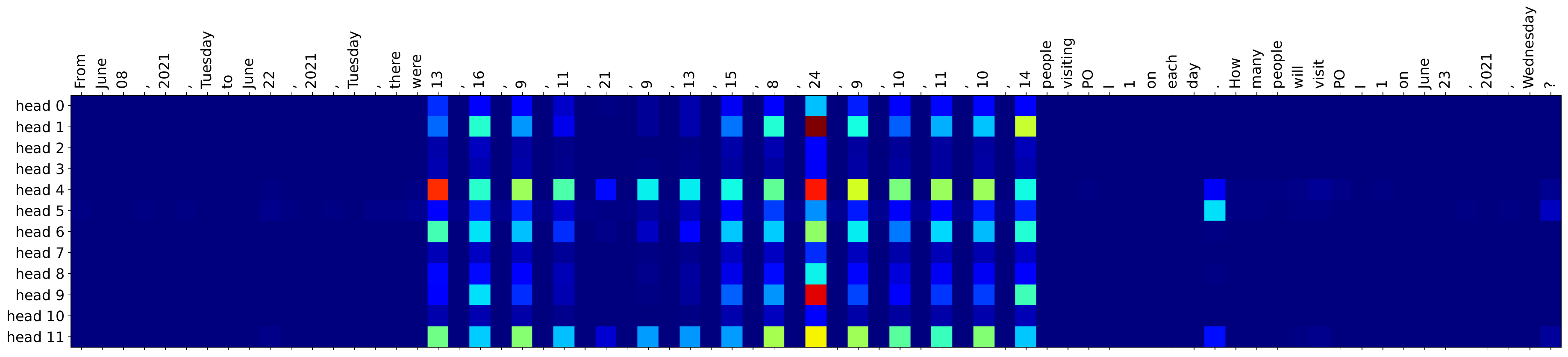}} \\
    \caption{Visualizations of attentions in the proposed \tname.}
    \label{fig:vis}
\end{figure*}

% To further investigate the why language models can work well for forecasting time series data under the proposed \tname\ setting, 
To validate the above hypothesis, we conduct two case studies and visualize the attentions (between the input sentence and the output sentence) learned by the language models (using the Bart model as an example in these case studies).
\begin{itemize}
    \item \textit{Case Study 1}
    \begin{itemize}
        \item \textbf{Input}: From September 06, 2019, Friday to September 20, 2019, Friday, the average temperature of region 1 was 65, 66, 70, 71, 71, 77, 78, 65, 70, 76, 74, 70, 64, 61, 64 degree on each day. What is the temperature going to be on September 21, 2019, Saturday?
        \item \textbf{Predicted}: The temperature will be 68 degree.
        \item \textbf{Ground Truth}: The temperature will be 71 degree.
    \end{itemize}
    \item \textit{Case Study 2}
    \begin{itemize}
        \item \textbf{Input}: From June 08, 2021, Tuesday to June 22, 2021, Tuesday, there were 13, 16, 9, 11, 21, 9, 13, 15, 8, 24, 9, 10, 11, 10, 14 people visiting POI 1 on each day. How many people will visit POI 1 on June 23, 2021, Wednesday?
        \item \textbf{Predicted}: There will be 11 visitors.
        \item \textbf{Ground Truth}: There will be 12 visitors.
    \end{itemize}
\end{itemize}
Specifically, in Figure~\ref{fig:vis}, we show the attention weights between the numerical predicted values in the generated outputs and all the tokens in the input prompts.
For each heatmap plot, with hotter regions representing larger attention values, the horizontal axis represents the input prompt (in token format) and the vertical axis represents different attention heads.

For Case 1 (Figure~\ref{fig:vis} (a)), we can clearly observe that head 6 pays more attention to the numerical sequential data in the input prompt while the other heads have higher attention to other semantic auxiliary information such as {\fontfamily{qcr}\selectfont September} and {\fontfamily{qcr}\selectfont Friday}. 
Moreover, head 7 shows attentions to both the input numerical historical observations (\eg, {\fontfamily{qcr}\selectfont 76}, {\fontfamily{qcr}\selectfont 74}, and {\fontfamily{qcr}\selectfont 64}) and semantic tokens simultaneously.
For Case 2, the attention visualization of two layers are displayed in Figure~\ref{fig:vis} (b) and Figure~\ref{fig:vis} (c).
It can be noticed that the learned attentions between the predicted value and the numerical historical values in the input prompt are mainly in layer 5 (Figure~\ref{fig:vis} (c)) of the Bart model.
Compared to layer 5, layer 4 (Figure~\ref{fig:vis} (b)) focuses on not only the numerical tokens but also contextual tokens.
More specifically, head 7 demonstrates large attention weights to semantic tokens like {\fontfamily{qcr}\selectfont there were} and {\fontfamily{qcr}\selectfont How many people will visit} whereas head 10 highlights the auxiliary date token {\fontfamily{qcr}\selectfont 22} as well as the punctuation token {\fontfamily{qcr}\selectfont ?}.
These case studies demonstrate that the attention learned by language models is able to take into account both the numerical value tokens (historical values) and auxiliary information tokens to make predictions under the \tname\ paradigm. 
% Based on the analysis of these two studies, we show that the language models can jointly learn the relation of numerical value tokens (historical values) at different time steps and the influence of the semantic tokens under the \tname\ paradigm.
It further justifies the rationale of leveraging language models for forecasting time series.

\section{Discussion and Conclusion}\label{sec:conclusion}
% In this paper, we present a novel task, \tname, which predicts time series using language models in a language generation manner. Given that this is the first work towards \tname\ task and no prior datasets are suitable, we build the first dataset aimed at investigating prompt-based forecasting. This large-scale dataset (\name) contains three real-world time series forecasting scenarios. To advance the research of \tname, we also develop a benchmark on the released dataset and provides a set of solid baselines including both state-of-the-art numerical forecasting methods and language generation models. The experimental results show that using language models under the \tname\ setting has good forecasting performance and generalization ability.
In this paper, we introduce a new task, \tname, which utilizes language models to predict time series in a language generation manner. As this is the first work on the \tname\ task, and no existing datasets are suitable, we construct the first dataset, \name, to investigate prompt-based forecasting. This large-scale dataset contains three real-world time series forecasting scenarios. To further advance the research of \tname, we also establish a benchmark on the released dataset and provide a set of strong baselines that include both state-of-the-art numerical forecasting methods and language generation models. The experimental results demonstrate that using language models in the \tname\ setting results in good forecasting performance and generalization ability.

\textbf{Broader Impact.}
% We believe that this study's findings would offer forward-thinking concepts and fresh insights for other researchers.
% The research of time series forecasting could contribute in solving problems such as climate change and resources allocation for social good.
We believe that the findings of this study will provide forward-thinking concepts and fresh insights for researchers working in the field of time series forecasting.
We also think that the proposed \tname\ task, as well as the \name\ dataset, could open new related research directions and provide visionary ideas about downstream applications that could be enabled by this work. Some potential directions for future research are discussed below. (1) \textit{Automatic Prompting}: In this paper, the transformation of numerical data to text is achieved through the use of templates. Although template-based prompting is efficient, it can be difficult to produce diverse prompts and fixed templates may introduce biases (biases towards certain templates). To address these limitations, a potential research direction is the development of automatic time series prompting or time series captioning (similar to image captioning~\cite{herdade2019image}), which utilizes generative models to describe time series data.
(2) \textit{Explainable \tname}:  
Another area that has yet to be fully investigated is the question of why models designed for language modeling tasks are able to predict time series. Recent research has begun to explore the use of language models for non-language downstream tasks~\cite{LIFT}. 
Although we have initially discussed our thoughts on why language models can be effective for time series forecasting under the \tname\ task through case studies, further studies on the interpretability and explainability of \tname\ models would be an interesting and valuable research direction to better understand the underlying mechanisms.
This can help us to design better models and understand the limitations of current models. 
By understanding the interpretability, we can also have a better understanding of how to apply these models to real-world problems.
(3) \textit{\tname\ QA and Chatbots}: 
The research of \tname\ task would promote the development of time series forecasting question-answering tasks and building chatbot applications with forecasting capabilities. Note that the \tname\ QA task differs from recent tasks such as TimeQA~\cite{timeqa}, which is proposed to answer general time-related questions based on Wikipedia text, and ForecastQA~\cite{forecastQA}, which is also based on text articles. The core of the \tname\ QA task is the question-answering ability for forecasting based on given sequential numerical value contexts.
This task can be applied to digital assistants such as Siri.
% The research of \tname\ would trigger and promote time series forecasting question answering tasks and building chatbots applications with forecasting ability.
% Note that the \tname\ QA task differs from the recent TimeQA~\cite{timeqa} that is proposed to answer general time related questions based on Wikipedia text and ForecastQA~\cite{forecastQA} which is also based on text articles. The core of \tname\ QA task is about the question-answering ability about forecasting upon the given sequential numerical value contexts.

\textbf{Limitations and Future Work.}
As the first attempt to create a dataset for the novel \tname\ task, in this dataset release, we have mainly focused on the univariate time series forecasting setting.
As discussed in Section~\ref{sec:mts}, one potential direction for our future work is to extend the dataset to include multivariate time series, which would support the research of more complex time series forecasting scenarios.
Another future work is to develop an auto-regressive method to further enhance the \tname\ performance on multi-step forecasting setting. For example, the language models could be fine-tuned to predict the value at the next time step. The generated sentence will then be recurrently appended to the input prompts to predict longer horizons.
Additionally, more challenging heterogeneous datasets that incorporate multiple data sources will also be considered for the \tname\ task in the future. 
The prompt-based paradigm can also be extended to other tasks such as time series classification and anomaly detection.
This would enable the proposed paradigm to be applied to a wider range of real-world scenarios.

\bibliographystyle{IEEEtran}
\bibliography{main}
\balance

\begin{IEEEbiography}[{\includegraphics[width=1in,height=1.25in,clip,keepaspectratio]{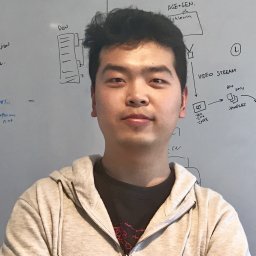}}]{Hao Xue}
is currently a Lecturer at the School of Computer Science and Engineering at UNSW Sydney. He earned his PhD from The University of Western Australia in 2020. After completing his doctorate, he worked as a Research Fellow at the School of Computing Technologies at RMIT University and UNSW Sydney. He was awarded the DAAD AInet Fellowship in 2022. His research interests include spatiotemporal data modelling, time series forecasting, language generation based forecasting, and data-efficient time series representation learning. 
\end{IEEEbiography}

\begin{IEEEbiography}[{\includegraphics[width=0.8in,height=1in,clip,keepaspectratio]{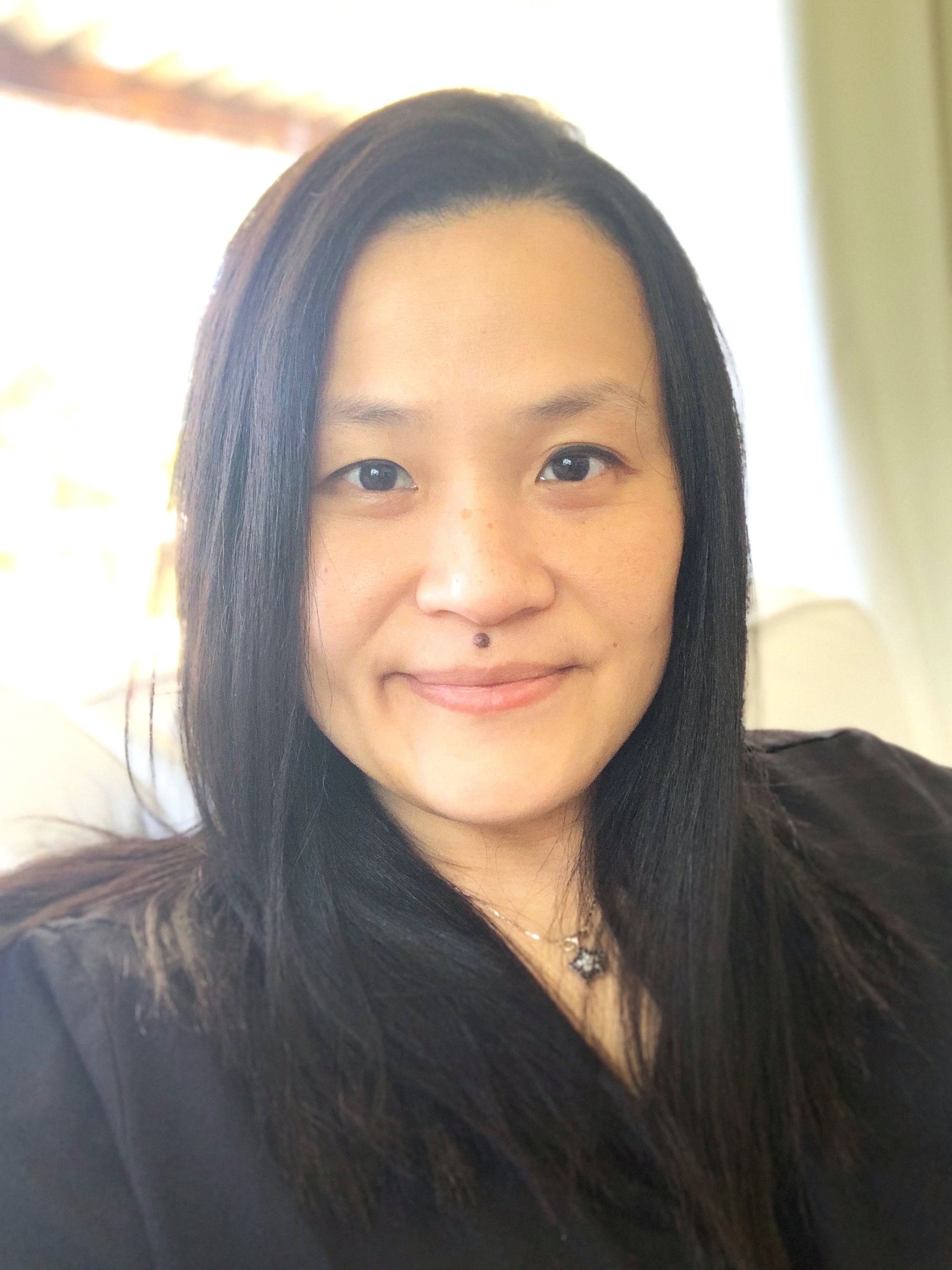}}]{Professor Flora Salim}
is the CISCO Chair of Digital Transport, School of Computer Science and Engineering, UNSW Sydney.Her research, on behaviour modelling, AI and machine learning on time-series and spatio-temporal sensor data, has been funded by the ARC, Humboldt Foundation, Bayer Foundation, Microsoft Research, Qatar National Research Fund, and many local and international industry partners. She won the Women in AI Awards 2022 ANZ Defence and Intelligence category.\end{IEEEbiography}

\newpage
\appendix

This Appendix document is organized as follows:
\begin{itemize}
    \item Section~\ref{appendix:cost} shows scatter plots representing the relationship between the cost and accuracy of each evaluated forecasting method.
    \item Section~\ref{appendix:host} introduces the hosting and licensing of our \name\ dataset.
    \item Section~\ref{appendix:datasheet} presents the \textit{Datasheet for Datasets} for our \name\ dataset.
\end{itemize}

\subsection{Additional Cost Analysis}\label{appendix:cost}

\begin{figure*}
  \centering

  \subfigure[CT sub-set]{
    \includegraphics[width=0.3\textwidth]{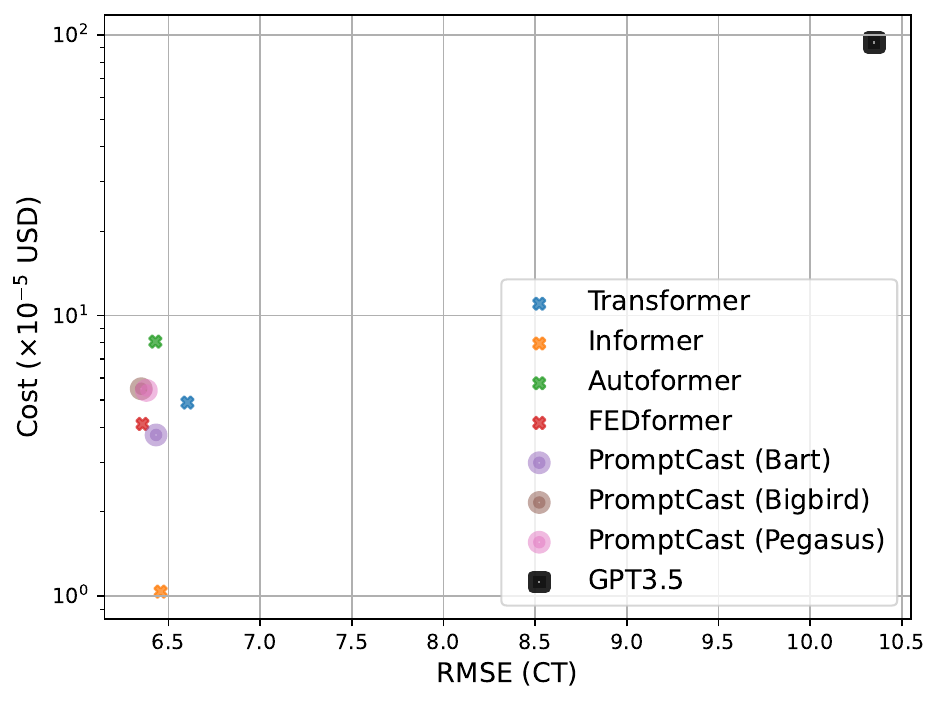}
    \label{fig:subfig1}
  }
  \subfigure[ECL sub-set]{
    \includegraphics[width=0.3\textwidth]{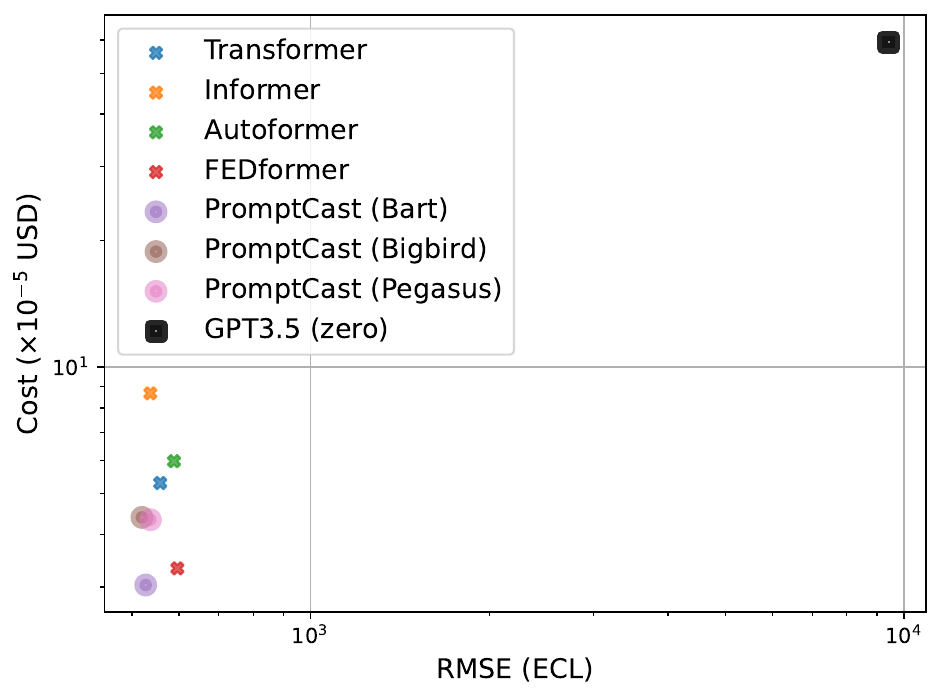}
    \label{fig:subfig2}
  }
  \subfigure[SG sub-set]{
    \includegraphics[width=0.3\textwidth]{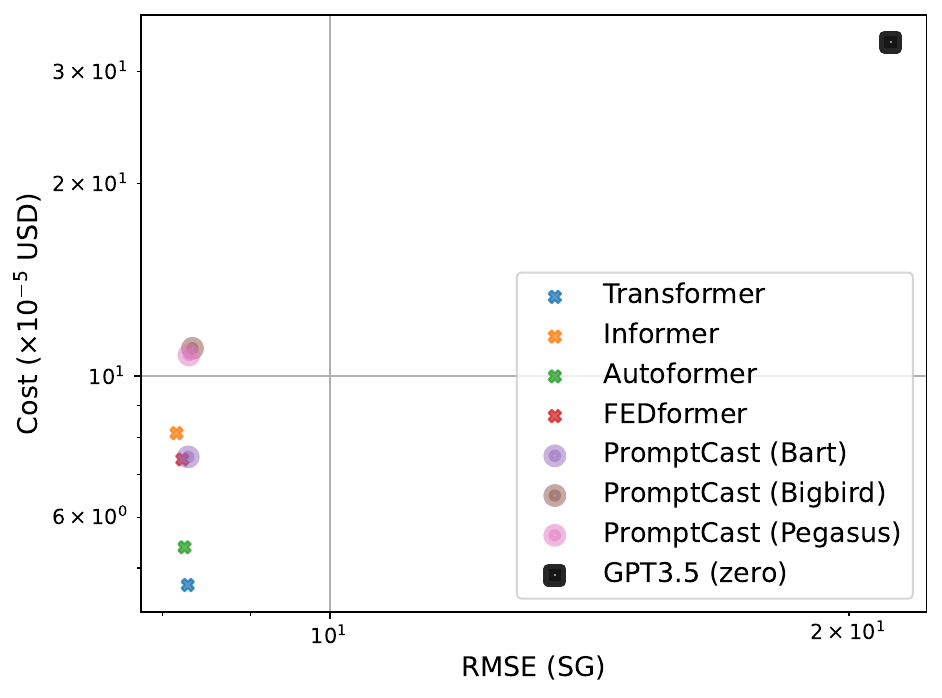}
    \label{fig:subfig3}
  }

  \caption{The plots of cost and RMSE accuracy on three sub-sets.}
  \label{fig:cost}
\end{figure*}

Three plots corresponding to the three sub-sets within PISA are presented in Figure~\ref{fig:cost}. These plots visually describe the cost and accuracy (\eg, RMSE) of different forecasting methods. As shown in the figure, our PromptCast presents a good forecasting accuracy while maintaining cost-effectiveness.

\subsection{\name\ Hosting and Licensing}\label{appendix:host}

\textbf{Source Data.}
\begin{itemize}
    \item CT: The source data for the CT sub-set can be accessed through the Average Daily Temperature Archive\footnote{https://academic.udayton.edu/kissock/http/Weather/default.htm}.
    Based on its description, this source data is available for research and non-commercial purposes only.
    \item ECL: The original data is available at UCI Machine Learning Repository\footnote{UCI Machine Learning Repository}. We used a processed version of the original data provided by Informer repository which is licensed under Apache License 2.0\footnote{https://github.com/zhouhaoyi/Informer2020/blob/main/LICENSE}.
    \item SG: We access the raw SafeGraph Weekly Patterns data through SafeGraph Data for Academics\footnote{https://www.safegraph.com/academics}. 
    ``\textit{SafeGraph\footnote{https://www.safegraph.com/}, a data company that aggregates anonymized location data from numerous applications in order to provide insights about physical places, via the SafeGraph Community. To enhance privacy, SafeGraph excludes census block group information if fewer than five devices visited an establishment in a month from a given census block group.}''
    According to the policy, it is against SafeGraph's terms of service to directly re-share raw SafeGraph data. However, it is acceptable under SafeGraph's terms of service to share aggregated and derived data, and to include data in chart and visual forms. We strongly recommend users to register SafeGraph Data for Academics before accessing our \name\ dataset.
\end{itemize}

\textbf{\name.}
The \name\ dataset and codes for models reported in the benchmark are available at \url{https://github.com/HaoUNSW/PISA}.
In this repository, we also provide some generated examples of language models for the \tname\ task. It's worth noting that only the validation sets are provided as \name\ examples in the above-mentioned repository during the submission period. After the acceptance decision notification, the full \name\ dataset (including train/val/testing sets) will be uploaded to the same repository and made publicly available.
The full dataset will also be available through HuggingFace Dataset\footnote{https://huggingface.co/datasets}, which will make it easier to use our dataset with HuggingFace models.

The \name\ dataset will be distributed under CC BY-NC-SA 4.0\footnote{https://creativecommons.org/licenses/by-nc-sa/4.0/}.

\subsection{\name\ Datasheet}\label{appendix:datasheet}

\noindent\textbf{Motivation}
\smallskip

\datasheet{For what purpose was the dataset created? (Was there a specific task in mind? Was there a specific gap that needed to be filled? Please provide a description.)}

The \name\ dataset is created to support the research of novel \tname\ task proposed in this paper.

\datasheet{Who created this dataset (e.g., which team, research group) and on behalf of which entity (e.g., company, institution, organization)?}

The authors (institution: School of Computer Science and Engineering, University of New South Wales, Sydney, Australia) of this paper created this \name\ dataset.

\datasheet{Who funded the creation of the dataset? (If there is an associated grant, please provide the name of the grant or and the grant name and number.)}

The creator of the dataset was supported by Australian Research Council (ARC) Discovery Project \textit{DP190101485}. 

\datasheet{Any other comments?}

None.

\noindent\textbf{Composition}
\smallskip

\datasheet{What do the instances that comprise the dataset represent (e.g., documents, photos, people, countries)? (Are there multiple types of instances (e.g., movies, users, and ratings; people and interactions between them; nodes and edges)? Please provide a description.)}

For the CT sub-set, the instances represent the daily temperature of a city.
For the ECL sub-set, the instances represent the daily electricity consumption of a user.
For the SG sub-set, the instances represent the daily visitor counts of a POI.

\datasheet{How many instances are there in total (of each type, if appropriate)?}

The proposed \name\ dataset includes 311,932 instances in total. The details of the partitions are presented in Table~\ref{tab:dataset}.

\datasheet{Does the dataset contain all possible instances or is it a sample (not necessarily random) of instances from a larger set? (If the dataset is a sample, then what is the larger set? Is the sample representative of the larger set (e.g., geographic coverage)? If so, please describe how this representativeness was validated/verified. If it is not representative of the larger set, please describe why not (e.g., to cover a more diverse range of instances, because instances were withheld or unavailable).)}

For the CT sub-set, 110 international cities are randomly selected to form the dataset. For the ECL sub-set,  We filtered users with missing values and randomly selected 50 users (from 321 users) with full records of the entire data collection period. For the SG sub-set, we randomly selected 324 POIs with full records. We list the data value range of each sub-set in Table~\ref{tab:dataset} and show the data distributions in Figure~\ref{fig:dis}.

\datasheet{What data does each instance consist of? (``Raw'' data (e.g., unprocessed text or images)or features? In either case, please provide a description.)}

Each instance consists of the input prompt and the output prompt. The input prompt is the input of a model and the output prompt is the desired output of the model. Our \name\ provides the input prompts and the corresponding output prompts in separate text files (\eg, val\_x\_prompt.txt and val\_y\_prompt.txt).

\datasheet{Is there a label or target associated with each instance? If so, please provide a description.}

Yes. As described in the above response, the output prompt is considered as the label.

\datasheet{Is any information missing from individual instances? (If so, please provide a description, explaining why this information is missing (e.g., because it was unavailable). This does not include intentionally removed information, but might include, e.g., redacted text.)}

There is no missing data (beyond what was intentionally omitted, \eg, the POI geo-location of SG sub-set).

\datasheet{Are relationships between individual instances made explicit (e.g., users' movie ratings, social network links)? ( If so, please describe how these relationships are made explicit.)}

Since the sliding window approach is applied (see Section~\ref{sec:raw_data}), the observation data of different instances might belong to the same object-of-interest. However, aach instance should be considered and treated as an independent instance with no relationship to other instances in \name. 

\datasheet{Are there recommended data splits (e.g., training, development / validation, testing)? (If so, please provide a description of these splits, explaining the rationale behind them.)}

Yes. Each sub-set in our \name\ is divided into train/val/test at the ratio of 7:1:2 by the chronological order. Please refer to Section~\ref{sec:raw_data} and Table~\ref{tab:dataset} for more details about the data splits.

\datasheet{Are there any errors, sources of noise, or redundancies in the dataset? (If so, please provide a description.)}

No.

\datasheet{Is the dataset self-contained, or does it link to or otherwise rely on external resources (e.g., websites, tweets, other datasets)? (If it links to or relies on external resources, a) are there guarantees that they will exist, and remain constant, over time; b) are there official archival versions of the complete dataset (i.e., including the external resources as they existed at the time the dataset was created); c) are there any restrictions (e.g., licenses, fees) associated with any of the external resources that might apply to a future user? Please provide descriptions of all external resources and any restrictions associated with them, as well as links or other access points, as appropriate.)}

The \name\ dataset is self-contained.

\datasheet{Does the dataset contain data that might be considered confidential (e.g., data that is protected by legal privilege or by doctor-patient confidentiality, data that includes the content of individuals' non-public communications)? (If so, please provide a description.)}

No.

\datasheet{Does the dataset contain data that, if viewed directly, might be offensive, insulting, threatening, or might otherwise cause anxiety? (If so, please describe why.)}

No.

\datasheet{Does the dataset relate to people? (If not, you may skip the remaining questions in this section.)}

No.

\datasheet{Does the dataset identify any subpopulations (e.g., by age, gender)? (If so, please describe how these subpopulations are identified and provide a description of their respective distributions within the dataset.)}

N/A.

\datasheet{Is it possible to identify individuals (i.e., one or more natural persons), either directly or indirectly (i.e., in combination with other data) from the dataset? (If so, please describe how.)}

N/A.

\datasheet{Does the dataset contain data that might be considered sensitive in any way (e.g., data that reveals racial or ethnic origins, sexual orientations, religious beliefs, political opinions or union memberships, or locations; financial or health data; biometric or genetic data; forms of government identification, such as social security numbers; criminal history)? (If so, please provide a description.)}

N/A.

\datasheet{Any other comments?}

None.

\noindent\textbf{Collection Process}
\smallskip

\datasheet{How was the data associated with each instance acquired? (Was the data directly observable (e.g., raw text, movie ratings), reported by subjects (e.g., survey responses), or indirectly inferred/derived from other data (e.g., part-of-speech tags, model-based guesses for age or language)? If data was reported by subjects or indirectly inferred/derived from other data, was the data validated/verified? If so, please describe how.)}

The data associated with each instance is acquired and derived from three data sources: CT, ECL, and SG. We have examined and verified the data. These data sources have also been used in the literature forecasting tasks.

\datasheet{What mechanisms or procedures were used to collect the data (e.g., hardware apparatus or sensor, manual human curation, software program, software API)? (How were these mechanisms or procedures validated?)}

N/A. Our \name\ dataset is based on existing data sources. We did not collect the data by ourselves.

\datasheet{If the dataset is a sample from a larger set, what was the sampling strategy (e.g., deterministic, probabilistic with specific sampling probabilities)?}

For the CT sub-set, 110 international cities are randomly selected to form the dataset. For the ECL sub-set,  We filtered users with missing values and randomly selected 50 users (from 321 users) with full records of the entire data collection period. For the SG sub-set, we randomly selected 324 POIs with full records.

\datasheet{Who was involved in the data collection process (e.g., students, crowdworkers, contractors) and how were they compensated (e.g., how much were crowdworkers paid)?}

All collection and annotation was done by the first author.

\datasheet{Over what timeframe was the data collected? (Does this timeframe match the creation timeframe of the data associated with the instances (e.g., recent crawl of old news articles)? If not, please describe the timeframe in which the data associated with the instances was created.)}

The data collection period of the data sources are reported in Table~\ref{tab:dataset}. The collection period of CT is 2017/01/01 - 2020/04/30.
The collection period of ECL is 2012/01/01 - 2014/12/31.
The collection period of SG is 2020/06/15 - 2021/09/05.

\datasheet{Were any ethical review processes conducted (e.g., by an institutional review board)? (If so, please provide a description of these review processes, including the outcomes, as well as a link or other access point to any supporting documentation.)}

N/A. The original data sources are publicly available and have been widely used in the literature.

\datasheet{Does the dataset relate to people? (If not, you may skip the remaining questions in this section.)}

No.

\datasheet{Did you collect the data from the individuals in question directly, or obtain it via third parties or other sources (e.g., websites)?}

N/A.

\datasheet{Were the individuals in question notified about the data collection? (If so, please describe (or show with screenshots or other information) how notice was provided, and provide a link or other access point to, or otherwise reproduce, the exact language of the notification itself.)}

N/A.

\datasheet{Did the individuals in question consent to the collection and use of their data? (If so, please describe (or show with screenshots or other information) how consent was requested and provided, and provide a link or other access point to, or otherwise reproduce, the exact language to which the individuals consented.)}

N/A.

\datasheet{If consent was obtained, were the consenting individuals provided with a mechanism to revoke their consent in the future or for certain uses? (If so, please provide a description, as well as a link or other access point to the mechanism (if appropriate).)}

N/A.

\datasheet{Has an analysis of the potential impact of the dataset and its use on data subjects (e.g., a data protection impact analysis) been conducted? (If so, please provide a description of this analysis, including the outcomes, as well as a link or other access point to any supporting documentation.)}

N/A.

\datasheet{Any other comments?}

None.

\noindent\textbf{Preprocessing/cleaning/labeling}
\smallskip

\datasheet{Was any preprocessing/cleaning/labeling of the data done (e.g., discretization or bucketing, tokenization, part-of-speech tagging, SIFT feature extraction, removal of instances, processing of missing values)? (If so, please provide a description. If not, you may skip the remainder of the questions in this section.)}

Yes. The details of the preprocessing/cleaning/labeling of the raw data sources are described in Section~\ref{sec:raw_data}. The prompting process of our \name\ dataset is provided in Section~\ref{sec:prompt_data}.

\datasheet{Was the "raw" data saved in addition to the preprocessed/cleaned/labeled data (e.g., to support unanticipated future uses)? (If so, please provide a link or other access point to the "raw" data.)}

For the CT and ECL sub-sets, the raw data sources are publicly available. For the SG sub-set, due to SafeGraph policy, the raw data will not be provided. Please refer to Section~\ref{appendix:host} for more details.

\datasheet{Is the software used to preprocess/clean/label the instances available? (If so, please provide a link or other access point.)}

Yes. In the GitHub repository (\url{https://github.com/HaoUNSW/PISA}), we provide the codes of transferring the source numerical data to prompts.

\datasheet{Any other comments?}

None.

\noindent\textbf{Uses}
\smallskip

\datasheet{Has the dataset been used for any tasks already? (If so, please provide a description.)}

No. This \name\ dataset is designed and introduced for the novel \tname\ task proposed in the paper.

\datasheet{Is there a repository that links to any or all papers or systems that use the dataset? (If so, please provide a link or other access point.)}

Considering that the proposed dataset \name\ is associated with the novel task \tname\ formulated in this paper, there is no repository for the time being. In the future, we do have a plan to create such a repository to summarize the papers related to this dataset or this task.

\datasheet{What (other) tasks could the dataset be used for?}

The \name\ dataset could also be used for the conventional numerical-based time series forecasting task.

\datasheet{Is there anything about the composition of the dataset or the way it was collected and preprocessed/cleaned/labeled that might impact future uses? (For example, is there anything that a future user might need to know to avoid uses that could result in unfair treatment of individuals or groups (e.g., stereotyping, quality of service issues) or other undesirable harms (e.g., financial harms, legal risks) If so, please provide a description. Is there anything a future user could do to mitigate these undesirable harms?)}

No.

\datasheet{Are there tasks for which the dataset should not be used? (If so, please provide a description.)}

No. We encourage other researchers to try our dataset on other related tasks.

\datasheet{Any other comments?}

None.

\noindent\textbf{Distribution}
\smallskip

\datasheet{Will the dataset be distributed to third parties outside of the entity (e.g., company, institution, organization) on behalf of
which the dataset was created? If so, please provide a description.}

Yes. During the reviewing period, the dataset (\ie, the validation sets as examples) is available \url{https://github.com/HaoUNSW/PISA}.
The full dataset will be publicly available on the same GitHub page for download by all interested third parties for research purpose after the acceptance decision notification.

\datasheet{How will the dataset will be distributed (e.g., tarball on website, API, GitHub) Does the dataset have a digital object identifier
(DOI)?}

The dataset will be distributed through the GitHub page and HuggingFace dataset page. Please refer to Section~\ref{appendix:host} for more details.

\datasheet{When will the dataset be distributed?}

The full dataset will be made publicly available after the acceptance decision notification.

\datasheet{Will the dataset be distributed under a copyright or other intellectual property (IP) license, and/or under applicable terms
of use (ToU)? If so, please describe this license and/or ToU, and provide a link or other access point to, or otherwise reproduce,
any relevant licensing terms or ToU, as well as any fees associated with these restrictions.}

Our \name\ dataset will be distributed under the CreativeCommons Attribution-NonCommercial-ShareAlike license (CC-BY-NC-SA). The terms of this license may be found at: \url{https://creativecommons.org/licenses/by-ncsa/2.0/legalcode}.

\datasheet{Have any third parties imposed IP-based or other restrictions on the data associated with the instances? If so, please
describe these restrictions, and provide a link or other access point to, or otherwise reproduce, any relevant licensing terms, as
well as any fees associated with these restrictions.}

No. There are no third party restrictions on the dataset.

\datasheet{Do any export controls or other regulatory restrictions apply to the dataset or to individual instances? If so, please
describe these restrictions, and provide a link or other access point to, or otherwise reproduce, any supporting documentation.}

No export controls or other regulatory restrictions apply to the dataset. However, we strongly recommend users to register SafeGraph Data for Academics before considering our \name\ dataset. Please refer to Section~\ref{appendix:host} for more details.

\datasheet{Any other comments?}

None.

\noindent\textbf{Maintenance}
\smallskip

\datasheet{Who will be supporting/hosting/maintaining the dataset?}

The authors of this paper will support/host/maintain the proposed \name\ dataset.

\datasheet{How can the owner/curator/manager of the dataset be contacted (e.g., email address)?}

The owner/curator/manager of the dataset can be contacted via: hao.xue1@unsw.edu.au.

\datasheet{Is there an erratum? If so, please provide a link or other access point.}

No erratum for the current version of \name\ dataset.

\datasheet{Will the dataset be updated (e.g., to correct labeling errors, add new instances, delete instances)? If so, please describe
how often, by whom, and how updates will be communicated to users (e.g., mailing list, GitHub)?}

The dataset will be updated in the future by the authors of this paper. The updates will focus on adding new forecasting scenarios, introducing more prompt templates, or expanding \name\ to multivarite time series setting. The updated will be communicated to users through the GitHub page and the HuggingFace page.

\datasheet{If the dataset relates to people, are there applicable limits on the retention of the data associated with the instances (e.g.,
were individuals in question told that their data would be retained for a fixed period of time and then deleted)? If so,
please describe these limits and explain how they will be enforced.}

N/A.

\datasheet{Will older versions of the dataset continue to be supported/ hosted/ maintained? If so, please describe how. If not, please
describe how its obsolescence will be communicated to users.}

Yes, all the versions of the dataset will be supported/ hosted/ maintained by the authors of this paper. 
The versioning information will be communicated to users through the GitHub page and the HuggingFace dataset page.

\datasheet{If others want to extend/augment/build on/contribute to the dataset, is there a mechanism for them to do so? If so, please
provide a description. Will these contributions be validated/verified? If so, please describe how. If not, why not? Is there a process
for communicating/distributing these contributions to other users? If so, please provide a description.}

If others want to extend/augment/build on/contribute to the dataset, the authors of this paper should be first contacted. 
Other researchers are encouraged to pull requests on the GitHub page. We will validate and verify the contributed data before merge the contributed data.

\datasheet{Any other comments?}

None.

\end{document}